\input harvmac
\input amssym.tex
\newcount\figno
\figno=0
\def\fig#1#2#3{
\par\begingroup\parindent=0pt\leftskip=1cm\rightskip=1cm\parindent=0pt
\global\advance\figno by 1
\midinsert
\epsfxsize=#3
\centerline{\epsfbox{#2}}
\vskip 12pt
{\bf Fig. \the\figno:} #1\par
\endinsert\endgroup\par
}
\def\figlabel#1{\xdef#1{\the\figno}}
\def\encadremath#1{\vbox{\hrule\hbox{\vrule\kern8pt\vbox{\kern8pt
\hbox{$\displaystyle #1$}\kern8pt}
\kern8pt\vrule}\hrule}}
\def\underarrow#1{\vbox{\ialign{##\crcr$\hfil\displaystyle
 {#1}\hfil$\crcr\noalign{\kern1pt\nointerlineskip}$\longrightarrow$\crcr}}}
%
\overfullrule=0pt

%
\def\tilde{\widetilde}
\def\bar{\overline}
\def\hat{\widehat}

\font\zfont = cmss10 
\font\litfont = cmr6

\def\bigone{\hbox{1\kern -.23em {\rm l}}}
\def\ZZ{\hbox{\zfont Z\kern-.4emZ}}
\def\half{{\litfont {1 \over 2}}}


\Title{hep-th/0504078} {\vbox{ \centerline{Two-Dimensional
Models}\smallskip
\bigskip\centerline{With $(0,2)$ Supersymmetry:  }
\bigskip\centerline{Perturbative Aspects    }}}
\smallskip
\centerline{Edward Witten}
\smallskip
\centerline{\it Institute For Advanced Study, Princeton NJ 08540 USA}

\medskip

\noindent
Certain perturbative aspects of two-dimensional sigma models with $(0,2)$ supersymmetry
are investigated. The main goal is to understand in physical terms how the mathematical theory of
``chiral differential operators'' is related to sigma models.  In the process, we obtain, for example,
an understanding of the one-loop beta function in terms of holomorphic data.
A companion paper will study nonperturbative behavior of these theories.

\Date{April, 2005}
\newsec{Introduction}

Two-dimensional sigma models with $(0,2)$ supersymmetry have
attracted much interest over the years, largely because, in the
conformally invariant case, they can be used to construct
supersymmetric compactifications of heterotic string theory.
These models have interesting nonperturbative effects that can
spoil conformal invariance even when it holds in perturbation
theory \ref\dsww{M. Dine, N. Seiberg, X.-G. Wen, and E. Witten,
``Nonperturbative Effects On The String World Sheet,'' Nucl. Phys.
{\bf B278} (1986) 769.}; much attention has focused on determining
conditions for exact conformal invariance (for example, see
\nref\dist{J. Distler and B. Greene, ``Aspects Of $(2,0)$ String
Compactifications,''
Nucl. Phys. {\bf B304} (1988) 1.}%
\nref\silv{E. Silverstein and E. Witten, ``Criteria For Conformal Invariance Of
$(0,2)$ Models,'' Nucl. Phys. {\bf B444} 161, hep-th/0503212.}%
\nref\seth{A. Basu and S. Sethi, ``World Sheet Stability Of $(0,2)$ Linear Sigma
Models,'' Phys. Rev. {\bf D68}:025003 (2003), hep-th/0303066.}%
\nref\beas{C. Beasley and E. Witten, ``Residues And World Sheet Instantons,''
JHEP 0310:065,2003, hep-th/0304115.}%
\refs{\dist - \beas}).
The present paper, however, will pursue a rather different direction.

By taking the cohomology of one of the supercharges of a (0,2)
model, one can construct a half-twisted model \ref\who{E. Witten,
``Mirror Manifolds And Topological Field Theory,'' in {\it Essays
On Mirror Manifolds}, ed. S.-T. Yau (International Press, 1992),
hep-th/9112056.} that is characterized by a chiral algebra.  This
chiral algebra contains much information about the dynamics of the
underlying model, but this has not yet been fully
 exploited. See
\ref\polyo{E. Witten, ``On The Landau-Ginzburg Description Of
$N=2$ Minimal Models,'' Int. J. Mod. Phys. {\bf A9} (1994) 4783,
hep-th/9304026.} for an example of determination of such a chiral
algebra.    The structure is further enriched because the elliptic
genus, a familiar invariant of two-dimensional supersymmetric
models, can be expressed in terms of supersymmetric physical
states which furnish a module for the chiral algebra.  (The
elliptic genus is defined using $(0,1)$ supersymmetry \ref\olyo{E.
Witten, ``Elliptic Genera And Quantum Field Theory,'' Commun.
Math. Phys. {\bf 109} (1987) 525, ``The Index Of The Dirac
Operator In Loop Space,'' in {\it Elliptic Curves And Modular
Forms In Algebraic Topology},'' ed. P. Landweber (Springer-Verlag,
1988).} and can be interpreted in terms of a module for a chiral
algebra when one has $(0,2)$ supersymmetry.) Correlation functions
of the half-twisted model are related in certain situations to
Yukawa couplings in heterotic string compactifications.  They have
been studied from many points of view; for recent discussion, see
\ref\katzsh{S. Katz and E. Sharpe, ``Notes On Certain (0,2)
Correlation Functions,'' hep-th/0406226; E. Sharpe, ``Notes on
Correlation Functions In (0,2) Theories,'' hep-th/0502064.}.

\nref\cdo{F. Malikov, V. Schechtman, and A. Vaintrob, ``Chiral De
Rham Complex,''
math.AG/9803041.}%
\nref\gdo{F. Malikov and V. Schechtman, ``Chiral de Rham Complex,
II,'' math.AG/9901065, ``Chiral Poincar\'e Duality,''
math.AG/9905008.}%
\nref\ngog{V. Gorbounov, F. Malikov, and V. Schechtman, ``Gerbes
Of Chiral Differential Operators,'' math.AG/9906117, ``Gerbes Of
Chiral Differential Operators, II,'' math.AG/0003170, ``Gerbes Of
Chiral Differential Operators,
III,'' math.AG/0005201.}%
In the present paper, we will study the half-twisted model {\it in
perturbation theory}. We will interpret the {\it perturbative}
approximation to the chiral algebra of the half-twisted model in
terms of the mathematical theory of ``Chiral Differential
Operators'' or CDO's \refs{\cdo-\ngog}.  This theory was also
developed independently in section 3.9 of \ref\bd{A. Beilinson and
V. Drinfeld, {\it Chiral Algebras} (American Mathematical Society,
2004).}; the CDO of a flag manifold had in essence been introduced
earlier \ref\ff{B. Feigin and E. Frenkel, ``A Family of
Representations of Affine Lie Algebras", Russ. Math. Surv. {\bf
43}, N 5 (1988) 221; ``Affine Kac-Moody Algebras and Semi-Infinite
Flag Manifolds", Comm. Math. Phys. {\bf 128} (1990) 161.}.  Some
of these developments have been motivated by potential
applications to the geometric Langlands program \ref\frenkel{E.
Frenkel, ``Affine algebras, Langlands duality and Bethe Ansatz,"
in {\it Proceedings of the International Congress of Mathematical
Physics},  ed. D. Iagolnitzer,  (International Press, 1995), p.
606, q-alg/9506003; ``Recent Advances In The Langlands Program,''
math.AG/0303074.}.
 Borrowing the mathematical results, we acquire a few
novel insights about the physics; for example, we interpret the
one-loop beta function in terms of holomorphic data. Perhaps our
results will also be of interest for mathematical applications of
CDO's.

Beyond perturbation theory, the chiral algebras of (0,2) models
are no longer related to CDO's.  What happens nonperturbatively
will be the subject of a separate paper, in which we will see that
instanton effects often change the picture radically, triggering
the spontaneous breaking of supersymmetry and making the chiral
algebra trivial.   This can only happen when the elliptic genus
vanishes, since a non-zero elliptic genus is an obstruction to
supersymmetry breaking.

If one drops the requirement of unitarity, theories obtained by
half-twisting a $(0,2)$ model are a special case of a larger class
of sigma models that lead to chiral algebras.  We describe this
larger class in section 2.  In section 3, we characterize the
chiral algebra that arises in perturbation theory from such a
model in terms of a sheaf of CDO's.  In section 4, we incorporate
unitarity and specialize to twisted versions of models with
$(0,2)$ supersymmetry.  In section 5, we perform explicit
calculations illustrating how the standard one-loop anomalies of
$(0,2)$ models -- the beta function and the chiral anomaly --
arise from the CDO point of view.  These examples are inspired by
and generalize an example considered in section 5.6 of \cdo\ as
well as the detailed analysis of anomalies in \ngog.

The question of how to interpret sheaves of CDO's in terms of
physics was originally raised by F. Malikov at the Caltech-USC
Center for Theoretical Physics in 1999-2000 and by E. Frenkel at a
conference on the geometric Langlands program held at the IAS in
the spring of 2004. These questions were the genesis of the
present paper. In addition, E. Frenkel generously explained a
number of relevant mathematical constructions, including chiral
algebras without stress tensors and \v{C}ech description of
operators and anomalies, some of which would have been quite
difficult to understand from the literature.  I am grateful for
his assistance. Finally, I would like to thank Meng-Chwan Tan for
a careful reading of the paper and spotting of many imprecisions.

\newsec{Chiral Algebras From Sigma Models}

\subsec{Classical Action}

In the present section, we consider the minimal structure of a
two-dimensional sigma model that enables one to define a chiral
algebra.  We consider a sigma model of maps $\Phi$ from a Riemann
surface $\Sigma$ to a complex manifold $X$. We pick a local
complex coordinate $z$ on $\Sigma$ to write formulas. The only
bosonic field in the model is $\Phi$ itself, which we describe
locally via fields $\phi^i(z,\bar z)$ (which are pullbacks to
$\Sigma$, via $\Phi$, of local complex coordinates on $X$). There
are also the following fermionic fields: $\rho^i$ is a
$(0,1)$-form on $\Sigma$ with values in $\Phi^*(TX)$; and
$\alpha^{\bar i}$ is a zero-form on $\Sigma $ with values in
$\Phi^*(\overline {TX})$.  Here $\Phi^*(TX)$ and
$\Phi^*(\overline{TX})$ are the pullback to $\Sigma$, via $X$, of
the holomorphic and antiholomorphic tangent bundles $TX$ and
$\overline{TX}$ of $X$.

We postulate a fermionic symmetry with \eqn\nurgo{\eqalign{ \delta
\phi^i&=\delta \alpha^{\bar i} = 0 \cr
                 \delta \bar\phi^{\bar i}& = \alpha^{\bar i}\cr
                  \delta\rho_{\bar z}{}^i& =-\partial_{\bar z} \phi^i.\cr}}
We call the generator of this symmetry $Q$. We introduce a $U(1)$
symmetry $R$ (which nonperturbatively may be violated by
instantons) under which $\alpha $ has charge 1, $\rho $ has charge
$-1$, and $\phi$ is invariant. So $Q$ has $R$-charge 1.

Clearly, $Q^2=0$ classically (and also quantum mechanically, as we
discuss later), and so any action of the form $I=\int |d^2z|\{Q,
V\}$ is $Q$-invariant, for any $V$. (Here $|d^2z|=idz\wedge d\bar
z$.) Choosing on $X$ a hermitian (not necessarily Kahler) metric
$ds^2=g_{i\bar j}d\phi^id\bar\phi^{\bar j}$, we take $V=-g_{i\bar
j}\rho^i\partial_z\bar\phi{}^{\bar j}$.  This leads to the action
that we use to define our most basic sigma model:
\eqn\nogno{I=\int |d^2z| \left(g_{i\bar j}\partial_{\bar
z}\phi^i\partial_z\bar \phi{}^{\bar j}+g_{i\bar j}\rho_{\bar
z}^i\partial_z\alpha^{\bar j}-g_{i\bar j,\bar k}\alpha^{\bar
k}\rho_{\bar z}^i\partial_z\bar\phi{}^{\bar j}\right).}

It is very natural to extend this model to include additional
fermionic fields valued in $\Phi^*(E)$, where $E$ is a holomorphic
vector bundle over $X$.  The generalization is important in
heterotic string compactification and hence is extensively
analyzed in the literature on $(0,2)$ models (for recent
discussion, see \katzsh); this more general possibility has also
been considered in the mathematical literature on CDO's. A special
case is that in which $E=TX$; the resulting model is then a
half-twisted version of the usual sigma model with $(2,2)$
supersymmetry, and the associated CDO has been called
mathematically the chiral de Rham complex (CDR). For brevity, we
will omit these generalizations. Other generalizations that have
been studied in the mathematical literature \nref\efr{E. Frenkel
and M. Szczesny, ``Chiral de Rham Complex and Orbifolds,''
math.AG/0307181.}%
\nref\lin{B. H. Lian and A. R. Linshaw, ``Chiral Equivariant
Cohomology, I,'' math.DG/0501084.}%
\refs{\efr,\lin} could plausibly be interpreted  physically in
terms of  the perturbative approximation to orbifolds and to
gauged sigma models, respectively, with $(2,2)$ supersymmetry.

Our goal in this section is to study, in perturbation theory, the
cohomology of $Q$ acting on local operators of this sigma model.
We will examine this more closely after describing some properties
of the quantum model, but first we make a few observations about
the classical theory. Classically, the model is conformally
invariant; the trace $T_{z\bar z}$ of the stress tensor vanishes.
The nonzero components of the stress tensor are $T_{zz}=g_{i\bar
j}\partial_z\phi^i\partial_z\bar\phi{}^{\bar j}$ and $T_{\bar
z\,\bar z}=g_{i\bar j}\left(\partial_{\bar z}\phi^i\partial_{\bar
z}\bar\phi{}^{\bar j}+ \rho^i D_{\bar z}\alpha^{\bar j}\right)$.
All components of the stress tensor are $Q$-invariant, but
crucially, $T_{\bar z\,\bar z}$ is trivial in cohomology, being
$\{Q,-g_{i\bar j}\rho^i\partial_{\bar z}\bar \phi{}^{\bar j}\}$.

We say that a local operator $\cal O$ inserted at the origin has
dimension $(n,m)$ if under a rescaling $z\to \lambda z$, $\bar
z\to \bar\lambda z$ (which is a symmetry of the classical theory),
it transforms as $\partial^{n+m}/\partial z^n\partial\bar z^m$,
that is, as $\lambda^{-n}\bar\lambda{}^{-m}$. Classical  local
operators have dimensions $(n,m)$ where $n$ and $m$ are
non-negative integers.\foot{Quantum mechanically, anomalous
dimensions shift the values of $n$ and $m$, but the difference
$n-m$ is unchanged.} But the cohomology of $Q$ vanishes except for
$m=0$. The reason for the last statement is that the rescaling of
$\bar z$ is generated by $\bar L_0=\oint d\bar z\, \bar z T_{\bar
z\bar z}$. As we noted in the last paragraph, $T_{\bar z\,\bar z}$
is of the form $\{Q,\dots\}$, so $\bar L_0=\{Q,W_0\} $ for some
$W_0$. Hence, if $\{Q, {\cal O}\}=0$, then $[\bar L_0,{\cal
O}]=\{Q,[W_0,{\cal O}]\}$.  If $[\bar L_0,{\cal O}]=m{\cal O} $
for $m\not= 0$, it follows that ${\cal O}$ is trivial in
cohomology.

By an argument similar to the above, if ${\cal O}$ is annihilated
by $Q$, then as an element of the cohomology, ${\cal O}(z)$ varies
holomorphically with $z$. Indeed,  $\partial_{\bar z}{\cal O}$ is
the commutator with ${\cal O}$ of $\bar L_{-1}=\oint d\bar
z\,T_{\bar z\bar z}$. We have $\bar L_{-1}=\{Q,W_{-1}\}$ for some
$W_{-1}$, and hence $\partial_{\bar z}({\cal O})=\{Q,[W_{-1},{\cal
O}]\}$.

We conclude with a few comments on the quantum theory. In section
2.3, we will show that quantum mechanically, just like
classically, $T_{\bar z\,\bar z}$ and $T_{z\bar z}$ are of the
form $\{Q,\dots\}$. Hence the same is true for the momentum
operator (called $\bar L_{-1}$ in the last paragraph) that
generates $\partial/\partial\bar z$. This implies that operators
${\cal O}(z)$ that represent cohomology classes vary
holomorphically with $z$, just as in the classical theory.  As for
the assertion that the cohomology of $Q$ on operators of dimension
$(n,m)$ is trivial for $m\not=0$, this statement is automatically
true quantum mechanically (at least in perturbation theory or when
quantum effects are small enough) if it is true classically, since
a vanishing cohomology continues to vanish after any small
perturbation.

Now, let ${\cal O}(z)$ and $\tilde{\cal O}(z')$ be two operators
that commute with $Q$, so that their product does so likewise.
Consider the operator product expansion or OPE of this product:
\eqn\considerit{{\cal O}(z)\tilde{\cal O}(z')\sim
\sum_{k}f_k(z-z'){\cal O}_k(z').} In general the coefficient
functions $f_k$ are not holomorphic.  But if we pass to the
cohomology and drop operators on the right hand side that are
$\{Q,\dots\}$, then the surviving coefficient functions are
holomorphic. In fact, $\partial/\partial\bar z$ acting on the left
hand side of \considerit\ gives terms that are cohomologically
trivial,\foot{Here we use the fact that $\{Q,{\cal O}\}=0$.  So
$\partial_{\bar z}{\cal O}=\{Q, S(z)\}$ for some $S(z)$, as we
argued before. Hence $\partial_{\bar z}{\cal O}(z)\cdot
\tilde{\cal O}(z')=\{Q,S(z)\tilde{\cal O}(z')\}$, where we use
also the fact that $[Q,\tilde{\cal O}]=0$.} so the $f_k$'s that
are not annihilated by $\partial/\partial \bar z$ multiply
operators ${\cal O}_k$ that are likewise cohomologically trivial.
We have established, roughly speaking, that the cohomology of $Q$
has a natural structure of a holomorphic chiral algebra, which we
will call ${\cal A}$.

We must warn the reader here of the following.  As in the
mathematical literature on this subject, the notion we use here of
a chiral algebra does not quite coincide with the usual physical
notion.  In fact, reparameterization invariance on the worldsheet
$\Sigma$ is not one of the axioms.  The sigma model \nogno\ is
generically not invariant at the quantum level under holomorphic
changes of coordinate on $\Sigma$ (because it is not invariant
under conformal rescalings of the metric of $\Sigma$, which such
changes of coordinate induce).  As we see later from various
points of view, such invariance is not necessarily recovered at
the level of the chiral algebra. Our operators ${\cal O}(z)$ vary
holomorphically in $z$, and have operator product expansions that
obey the usual relations of holomorphy, associativity, and
invariance under translation and rescaling of $z$, but not
necessarily  invariance under arbitrary holomorphic
reparameterization of $z$. Our chiral algebras are in general only
defined locally, requiring a choice of complex parameter $z$ up to
translations and scaling, or alternatively, requiring a flat
metric up to scaling on the Riemann surface $\Sigma$.  This is
enough to define a chiral algebra on a surface of genus one, but
to define the chiral algebra on a Riemann surface of higher genus
requires more analysis, and is potentially obstructed by an
anomaly involving $c_1(\Sigma)$ and $c_1(X)$ that we will meet in
sections 2.3 and 3.5.

\bigskip\noindent{\it Relation To The Elliptic Genus}

Though in this paper we focus on operators, it is also possible to
construct states by canonical quantization of the theory on
$\Bbb{R}\times {\bf S}^1$.  The $Q$-cohomology of such states
furnishes a module ${\cal V}$ for the chiral algebra ${\cal A}$
that we obtained from the $Q$-cohomology of operators.  In case
$X$ is a Calabi-Yau manifold, the usual operator-state
correspondence gives a natural isomorphism from operators to
states. In that case, therefore, ${\cal V}$ is isomorphic to
${\cal A}$ itself. That is not so in general.

By counting bosonic and fermionic states in ${\cal V}$ of energy
$n$ one can form a modular function called the elliptic genus
which  has no quantum corrections,\foot{Absence of quantum
corrections is proved using the fact that both the energy and the
operators $(-1)^F$ that distinguishes bosonic and fermionic states
are exactly conserved quantum mechanically.} making it effectively
computable. Explicitly, the elliptic genus is
$V(q)=q^{-d/12}\sum_{n=0}^\infty q^n\,\Tr_{{\cal V}_n}{}(-1)^F $,
where ${\cal V}_n$ is the space of supersymmetric states of energy
$n$ and $\Tr_{{\cal V}_n}{}(-1)^F$ is its ``Euler characteristic''
(difference of bosonic and fermionic dimensions). When the
elliptic genus is nonzero, ${\cal V}$ is nonempty and hence
supersymmetry is not spontaneously broken. One can form an
analogous generating series for ${\cal A}$, with the operators
graded by dimension, and at least in perturbation theory this
function appears to have modular properties (though it is not
clear that this statement has a natural path integral proof).
Explicitly, this series, considered at the perturbative level in
section 5.6 of \cdo, is
$A(q)=q^{-d/12}\sum_{n=0}^\infty\,q^n\,\Tr_{{\cal A}_n}{}(-1)^F$,
with ${\cal A}_n$ being the space of operators of dimension $n$.
 However, this function, though constant in perturbation theory,
 {\it does} have nonperturbative quantum corrections (except on Calabi-Yau manifolds),
 because instanton corrections do not preserve the grading by dimension of
 an operator.  So even when $A(q)$ is nonzero in perturbation theory,
 it may vanish nonperturbatively.  In fact, $\Bbb{CP}^1$ gives an
 example, as we will discuss elsewhere.

\subsec{Moduli Of The Chiral Algebra}

Here we will consider the moduli of the chiral algebra found in
the last section.

There are a few obvious considerations.  The chiral algebra does
not depend on the hermitian metric $g_{i\bar j}$ that was used in
writing the classical action, since this metric appears in the
action entirely inside a term of the form $\{Q,\dots\}$.

The chiral algebra does depend on the complex structure of $X$,
because this enters in the definition of the fields and the
fermionic symmetry. In fact, the chiral algebra varies
holomorphically with the complex structure of $X$.  It is possible
to show this by showing that if $J$ denotes the complex structure
of $X$, then an antiholomorphic derivative $\partial/\partial \bar
J$ changes the action by terms of the form $\{Q,\dots\}$.

If $B$ is a closed two-form on $X$, then we can add to the action
a topological invariant \eqn\yurko{I_B=\int_\Sigma\Phi^*(B).}
Being a topological invariant, $I_B$ is invariant under any local
deformation of the fields and in particular under $Q$. Including
this term in the action has the effect of introducing a factor
$\exp(-I_B)$ in the path integral.  In perturbation theory, we
consider only degree zero maps $\Phi:\Sigma\to B$, so $I_B=0$ and
this factor equals 1.  Hence, the interaction $I_B$ is really not
relevant for the present paper. Nonperturbatively, $I_B$ affects
the weights assigned to instantons and can affect the chiral
algebra.

Our focus here is on a more subtle possibility involving a
topologically non-trivial $B$-field which is {\it not} closed and
so does affect perturbation theory. First, we describe the
situation locally.  Let $T={1\over 2}T_{ij}d\phi^i\wedge d\phi^j$
be any two-form on $X$ that is of type $(2,0)$.   Let
\eqn\hoxo{I_T=-\int |d^2z| \{Q,T_{ij}\rho^i\partial_z\phi^j\}} or
in more detail \eqn\ujjo{I_T =\int
|d^2z|\left(T_{ij}\partial_{\bar z}\phi^i\partial_{ z}\phi^j
-T_{ij,\bar k}\alpha^{\bar k}\rho^i\partial_z\phi^j\right).} Here
$T_{ij,\bar k}=\partial T_{ij}/\partial\bar \phi{}^{\bar k}$. As
will become clear, $T$ is best understood as a two-form gauge
field, like the object $B$ considered in the last paragraph,
except that  $B$ was constrained to be closed but not necessarily
of any particular Hodge type, while $T$ is of type $(2,0)$ but not
necessarily closed.

As written, $I_T$ is $Q$-trivial and depends on the choice of a
$(2,0)$-form $T$.  But in fact, with some mild restrictions, the
definition of $I_T$ depends only on the three-form ${\cal H}=dT$,
and makes sense even if $T$ is not globally-defined as a two-form
(and must be interpreted as a two-form gauge field, in the sense
of string theory, or in terms of mathematical theories such as
connections on gerbes or Cheeger-Simons differential characters).

In brief, we will find that $I_T$ can be defined for any closed
form ${\cal H}$ that is of type $(3,0)\oplus (2,1)$, but the
formula \hoxo\ that expresses $I_T$ as $\{Q,\dots\}$ is only valid
globally if it is true globally that ${\cal H}=dT$ for some $T$ of
type (2,0). Hence, the chiral algebra ${\cal A}$ depends on the
cohomology class (in a certain sense that will be clarified later)
of ${\cal H}$.

In fact, as $T$ is of type $(2,0)$, ${\cal H}=dT$ is a sum of
terms of types $(3,0)$ and $(2,1)$. The second term in \ujjo\ is
already written in terms of ${\cal H}$, since $T_{ij,\bar k}$ is
simply the $(2,1)$ part of ${\cal H}$. The first term in \ujjo\
can likewise be expressed in terms of ${\cal H}$. Recalling that
$|d^2z|=idz\wedge d\bar z$, we write that first term (which is a
generalization of the Wess-Zumino anomaly functional \ref\wz{J.
Wess and B. Zumino, ``Consequences Of Anomalous Ward Identities,''
Phys. Lett. {\bf 37B} (1971) 95.}) as
\eqn\nujjo{I_T^{(1)}=-{i\over 2} \int_\Sigma\,T_{ij}d\phi^i\wedge
d\phi^j=-i\int_\Sigma\Phi^*(T).} Suppose now that $C$ is a
three-manifold whose boundary is $\Sigma$ and over which the map
$\Phi:\Sigma\to X$ extends. Then, if $T$ is globally defined as a
$(2,0)$-form, the relation ${\cal H}=dT$ implies, via Stokes'
theorem, that \eqn\ucno{I_T^{(1)} =-i\int_C\Phi^*({\cal H}),} a
formula which expresses $I_T^{(1)}$, and hence $I_T$, just in
terms of ${\cal H}$.

We do not actually want to limit ourselves to the case that $T$ is
globally-defined as a two-form.  In general, the right hand side
of \ucno\ makes sense as long as ${\cal H}$ is globally defined.
But it depends on the choice of $C$ and of the extension of
$\Phi$.  To make this explicit, we denote it as $I_T^{(1)}(C)$. If
$C$ and $C'$ are two three-manifolds with boundary $\Sigma$ and
chosen extensions of $\Phi$, they glue together (after reversing
the orientation of $C'$) to a closed three-manifold $D$ with a map
$\Phi:D\to X$. Then
$I_T^{(1)}(C)-I_T^{(2)}(C')=-i\int_D\Phi^*({\cal H})$. In quantum
theory, a shift in the (Euclidean) action by an integral multiple
of $2\pi i$ is irrelevant; the indeterminacy in $I_T^{(1)}$ is of
this nature if ${\cal H}/2\pi$ has integral periods.
Nonperturbatively, for the quantum theory associated with the
classical actions considered here to be well-defined, ${\cal H}$
must obey this condition, and in particular, the integrality means
that continuous moduli of ${\cal H}$ that are present in
perturbation theory may be absent in the nonperturbative theory.
In the present paper, as we consider only perturbation theory, we
will not see this effect.

In writing $I_T^{(1)}$ in terms of ${\cal H}$, we have assumed
that $\Phi$ extends over {\it some} three-manifold $C$ of boundary
$\Sigma$. This assumption is certainly valid in perturbation
theory, in which one considers topologically trivial maps $\Phi$;
such maps extend over any chosen $C$.  Nonperturbatively, this
framework for defining $I_T^{(1)}$ is not adequate, as $C$ may not
exist. To define $I_T^{(1)}$ in general, one must interpret $T$ as
a two-form gauge field (or in terms of gerbes or differential
characters); in that framework, if ${\cal H}/2\pi$ is associated
with an intetgral cohomology class, $I_T^{(1)}$ is always
naturally defined mod $2\pi i$. In general, $T$ is not completely
determined as a two-form gauge field by its curvature ${\cal
H}=dT$, as one may add a flat two-form gauge field to $T$. Hence,
the functional $I_T^{(1)}$ does depend in general on global
information not contained in ${\cal H}$, but this dependence does
not affect perturbation theory, which only depends on ${\cal H}$.
This is why the sheaf of CDO's, as defined in the mathematical
literature \refs{\cdo,\bd}, and as interpreted in this paper from
a physical viewpoint, only depends on ${\cal H}$.

\bigskip\noindent{\it Moduli }

So far, we have locally a two-form gauge field $T$ that is of type
$(2,0)$, and whose curvature ${\cal H}=dT$ is hence of type
$(3,0)\oplus (2,1)$. Actually, given any closed form ${\cal H}$ of
type $(3,0)\oplus (2,1)$, we can always find locally a
$(2,0)$-form $T$ with ${\cal H}=dT$. To see this, we first select
{\it any} local two-form $W$ with ${\cal H}=dW$. $W$ exists
because of the Poincar\'e lemma.  {\it A priori}, $W$ is a sum of
terms $W^{(2,0)}+W^{(1,1)}+W^{(0,2)}$ of the indicated types.  Now
as $\bar\partial W^{(0,2)}=0$ (since ${\cal H}$ has no component
of type $(0,3)$), the $\bar\partial$ version of the Poincar\'e
lemma asserts that locally $W^{(0,2)}=\bar\partial\lambda$ for
some $(0,1)$-form $\lambda$. Upon replacing $W$ by $\tilde
W=W-d\lambda$, we have a new two-form $\tilde W$ with ${\cal
H}=d\tilde W$, and such that $\tilde W$ has a decomposition
$\tilde W=\tilde W^{(2,0)}+\tilde W^{(1,1)}$. Now again as
$\bar\partial\tilde W^{(1,1)}=0$, we have locally $\tilde
W^{(1,1)}=\bar\partial\tilde\lambda$. Finally, $T=\tilde
W-d\tilde\lambda$ is the desired $(2,0)$-form with ${\cal H}=dT$.

Hence, given any closed form ${\cal H}$ of type
$(3,0)\oplus(2,1)$, we can locally represent it as ${\cal H}=dT$
for some $(2,0)$-form $T$ and write the $Q$-invariant functional
$I_T$. The definition of $I_T$ essentially depends only on ${\cal
H}$ (modulo terms that do not affect perturbation theory), but the
formula \hoxo, which locally shows that $I_T$ is $Q$-trivial,
makes sense globally only if $T$ is globally-defined as a
two-form.  So the moduli, in perturbation theory, of the chiral
algebra derived from the sigma model with target a complex
manifold $X$ are parameterized by the closed form ${\cal H}$ of
type $(3,0)\oplus (2,1)$, modulo forms that can be written
globally as $dT$ for $T$ of type $(2,0)$.   Nonperturbatively, and
thus beyond the scope of this paper, the situation is somewhat
different, as ${\cal H}$ must be an integral class, and the flat
$B$-fields will also enter.

\bigskip\noindent{\it Interpretation Via $H^1(X,\Omega^{2,cl})$}

Now we want to describe more precisely what sort of cohomology
class ${\cal H}$ represents.

 Let $U_a$, $a=1,\dots,s$ be a collection of small open sets
giving a good cover of $X$. (This means that the individual $U_a$
and all of their intersections are open balls.)

Suppose that (as in our problem) we are given a closed form ${\cal
H}$ on $X$ that is of type $(3,0)\oplus(2,1)$. As we have seen, in
each $U_a$, there is a $(2,0)$-form $T_{a}$ with ${\cal
H}=dT_{a}$. On each intersection $U_{ab}=U_a\cap U_b$, let
$T_{ab}=T_a-T_b$. Clearly, \eqn\longo{T_{ab}=-T_{ba}} for each
$a,b$, and \eqn\hombo{T_{ab}+T_{bc}+T_{ca}=0} for each $a,b,c$.
Moreover, $dT_{ab}=0$ and hence (as $T_{ab}$ is of definite type
$(2,0)$) $\partial T_{ab}=\bar\partial T_{ab}=0$. The $T_{ab}$ are
not uniquely determined by ${\cal H}$.  The definition of $T_{a}$
would allow us to shift $T_a\to T_a+S_a$, where $S_a$ is a
$(2,0)$-form on $U_ a$ obeying $dS_a=0$ (and hence $\partial
S_a=\bar\partial S_a=0$). This change in the $T_a$ induces the
equivalence \eqn\honon{T_{ab}\to T_{ ab}+S_a-S_b.}

One defines $\Omega^2(X)$ as the sheaf of  $(2,0)$-forms on $X$,
and $\Omega^{2,cl}(X)$ as the sheaf of such forms that are
annihilated by $\partial$. (The label ``$cl$'' is short for
``closed'' and refers to forms that are closed in the sense of
being annihilated by $\partial$.)  A section of $\Omega^{2,cl}(X)$
that is holomorphic in a given set $U\subset X$ is a $(2,0)$-form
in $U$ that is annihilated by both $\bar\partial$ and  $\partial$.
Similarly, $\Omega^{n,cl}$ is the sheaf whose sections are $(n,0)$
forms that are annihilated by $\partial$; its holomorphic sections
are also annihilated by $\bar\partial$.   In the last paragraph,
we found in each double intersection $U_a\cap U_b$ a holomorphic
section $T_{ab}$ of $\Omega^{2,cl}(X)$.  The identities \longo\
and \hombo\ and equivalence \honon\ means that these fit together
into an element of the \v{C}ech cohomology group
$H^1(X,\Omega^{2,cl}(X))$.

If ${\cal H}$ is globally of the form $dT$ for some $(2,0)$-form
$T$, then we can take all $T_a$ to equal $T$, whereupon all
$T_{ab}$ vanish.  So we have obtained a map from the space of
closed forms ${\cal H}$ of type $(3,0)\oplus (2,1)$, modulo forms
that are globally $dT$ for $T$ of type $(2,0)$, to
$H^1(X,\Omega^{2,cl})$.

Conversely, suppose that we are given an element of this sheaf
cohomology group, represented by such a family $T_{ab}$. Let $f_a$
be a partition of unity subordinate to the cover $U_a$ of $X$.
This means that the $f_a$ are continuous functions on $X$ that
vanish outside $U_a$ and obey $\sum_a f_a=1$. Let ${\cal H}_a$ be
the three-form defined in $U_a$ by ${\cal H}_a=\sum_c df_c \wedge
T_{ac}$. ${\cal H}_a$ is well-defined throughout $U_a$, since in
$U_a$, $df_c$ vanishes wherever $T_{ac}$ is not defined.
Obviously, ${\cal H}_a$ is of type $(3,0)\oplus(2,1)$, since the
$T_{ac}$ are of type $(2,0)$, and moreover $d{\cal H}_a=0$. For
any $a$ and $b$, we have ${\cal H}_a-{\cal H}_b=\sum_c df_c\wedge
(T_{ac} -T_{bc})$. Using \hombo, this is $\sum_c df_c \wedge
T_{ab}=d\left(\sum_c f_c\right)\wedge T_{ab}$. This vanishes, as
$\sum_c f_c=1$.  So for all $a,b$, ${\cal H}_a={\cal H}_b$ on
$U_a\cap U_b$. The ${\cal H}_a$ thus fit together to a closed
three-form ${\cal H}$ that is of type $(3,0)\oplus (2,1)$. Thus,
we have found a map from the \v{C}ech cohomology group
$H^1(X,\Omega^{2,cl}(X))$ to closed forms ${\cal H}$ of type
$(3,0)\oplus (2,1)$, modulo those that are globally $dT$ for $T$
of type $(2,0)$. We leave it to the reader to verify that the two
maps we have described are inverses.

The conclusion of this analysis is that the sigma models
considered here, and therefore the chiral algebras derived from
them, can be ``shifted'' in a natural way by an interaction $I_T$
determined by an element ${\cal H}$  of $H^1(X,\Omega^{2,cl}(X))$.

Such an ${\cal H}$, being a closed three-form, also determines a
class in the de Rham cohomology $H^3(X,\Bbb{R})$, but ${\cal H}$
may vanish in de Rham cohomology even if it is nonzero in
$H^1(X,\Omega^{2,cl})$.  This occurs if ${\cal H}$ can be globally
written as $dT$ for some two-form $T$, but $T$ cannot be globally
chosen to be of type $(2,0)$.  For an example, let
$X=\Bbb{C}^2\times E$, where $\Bbb{C}^2$ has complex coordinates
$x,y$ and the elliptic curve $E$ is the quotient of the complex
$z$ plane by a lattice.  Let ${\cal H}=dx\wedge dy\wedge d\bar z$.
Then  ${\cal H}=d(x\,dy\wedge d\bar z)$, but ${\cal H}$ cannot be
written as $dT$ with $T$ of type $(2,0)$.

\subsec{Anomalies}

Before investigating the quantum properties of the model, the most
basic question to consider is whether it exists at all -- whether
the Lagrangian that we have written leads to some kind of quantum
theory.

A failure of the model to exist, even in perturbation theory,
would come from an anomaly in the path integral of the
world-volume fermions $\alpha^{\bar i}$ and $\rho^j$.  (For such
sigma model anomalies, see \ref\sigmamodel{G. W. Moore and P.
Nelson, ``The Etiology Of Sigma Model Anomalies,'' Commun. Math.
Phys. {\bf 100} (1985) 83}.) In this discussion, we can omit the
interaction $I_T$, as anomalies do not depend on continuously
variable couplings such as this one. Considering only the basic
action \nogno, the kinetic energy for the fermions is $(
\rho,D\alpha)=\int |d^2z|\, g_{i\bar i}\rho^iD\alpha^{\bar i}$,
where $D$ is the $\partial$ operator on $\Sigma $ acting on
sections of $ \Phi^*(\overline{ TX})$. Equivalently, if we pick a
spin structure on $\Sigma$, $D$ can be interpreted as a Dirac
operator on $\Sigma$ acting on sections of $V=\bar K^{-1/2}\otimes
\Phi^*(\overline {TX})$, with $K$ the canonical bundle of $\Sigma
$ and $\bar K$ its complex conjugate.

We consider a family of maps $\Phi:\Sigma\to X$, parameterized by
a base $B$.   In fact, in the path integral we want to consider
the universal family of all maps of $\Sigma$ to $X$.  These maps
fit together into a map $\Phi:\Sigma\times B\to X$.  To make sense
of the quantum path integral, we must be able to interpret the
determinant of $D$ as a function on $B$, but mathematically, it is
interpreted in general as a section of a determinant line bundle
${\cal L}$.   The quantum theory can only exist if ${\cal L}$ is
trivial. Conversely, the quantum theory will exist if ${\cal L}$
is trivial and can be trivialized by a local formula similar to
the Green-Schwarz anomaly cancellation mechanism.

The theory of  determinant line bundles is usually expressed in
terms of a family of $\bar D$ operators, while here we have a
family of $D$ operators. The $D$ operators would be converted into
$\bar D$ operators if we reverse the complex structure on
$\Sigma$, but in most of this paper, the formulas look much more
natural with the complex structure as we have chosen it.  (If we
were to reverse the complex structure, the $D$ operator of the
fermions would become a $\bar D$ operator, but our chiral algebra
would be antiholomorphic.)   At any rate, the theory of
determinants of $D$ operators is isomorphic to the theory of
determinants of $\bar D$ operators, so we can borrow the usual
results.

\nref\as{M. F. Atiyah and I. M. Singer, ``Dirac Operators Coupled
To Vector Potentials,'' Proc. Nat. Acad. Sci. {\bf 81} (1984)
2597.}%
\nref\freed{J. M. Bismut and D. Freed, ``The Analysis Of Elliptic
Families I: Metrics And Connections On Determinant Bundles,''
Commun. Math. Phys. {\bf 106} (1986) 59.}%

 The basic obstruction to triviality of ${\cal L}$ is its
first Chern class.  By the family index theorem, applied to
anomalies in \refs{\as,\freed}, the first Chern class of ${\cal
L}$ is $\pi_*({\rm ch}_4(V))$, where ${\rm ch}_4$ is the dimension
four part of the Chern character, and $\pi:\Sigma\times B\to B$ is
the projection to the second factor. This vanishes if  ${\rm
ch}_4(V)$ vanishes before being pushed down to $B$.\foot{If ${\rm
ch}_4(V)\not=0$ but $\pi_*({\rm ch}_4(V))=0$, then ${\cal L}$ is
trivial but cannot be trivialized by a local Green-Schwarz
mechanism, so the quantum sigma model does not exist.} To evaluate
this, we note that ${\rm ch}_4(\overline{TX})=p_1(X)/2$ and that
tensoring with $\bar K^{-1/2}$ adds an additional term
$c_1(\Sigma)c_1(X)/2$. Here $p_1(X)$ is the first Pontryagin class
of the ordinary, real tangent bundle of $X$; there is a natural
way to divide it by 2 to get an integral characteristic class. The
condition for vanishing is thus that \eqn\invi{0={1\over
2}c_1(\Sigma)c_1(X)={1\over 2}p_1(X).} The first condition means
at the level of differential forms that either $c_1(X)=0$ and
$\Sigma$ is arbitrary, or $c_1(X)\not= 0$, and we must restrict
ourselves to Riemann surfaces $\Sigma$ with $c_1(\Sigma)=0$.

The characteristic class $p_1(X)/2$  can be interpreted as an
element of $H^2(X,\Omega^{2,cl})$.\foot{This statement is possibly
most familiar for Kahler manifolds, where $p_1(X)$ is represented
by a form of type $(2,2)$, annihilated by both $\bar\partial$ and
$\partial$ and thus representing an element of
$H^2(X,\Omega^{2,cl})$. However, on any complex manifold, $p_1(X)$
can be represented by a closed form of type $(2,2)\oplus
(3,1)\oplus (4,0)$.  To see this, pick any connection on the
holomorphic tangent bundle $TX$ whose $(0,1)$ part is the natural
$\bar\partial$ operator of this bundle.  Since $\bar\partial^2=0$,
the curvature of such a connection is of type $(2,0)\oplus (1,1)$,
as a result of which, for every $k\geq 0$, $c_k(TX)$ (which is a
polynomial of degree $k$ in the curvature, and is usually
abbreviated as $c_k(X)$) is described by a form of type
$(k,k)\oplus (k+1,k-1)\oplus \dots \oplus (2k,0)$, and represents
an element of $H^k(X,\Omega^{k,cl}(X))$. In particular, $c_1(X)$
represents an element of $H^1(X,\Omega^{1,cl}(X))$, and
$p_1=2c_2(X)-c_1^2(X)$ represents an element of
$H^2(X,\Omega^{2,cl}(X))$.} We will meet it in this guise in
sections  3.5 and 5.2. Likewise, as explained in the footnote,
$c_1(X)$ corresponds to an element of $H^1(X,\Omega^{1,cl})$ and
$c_1(\Sigma)$ to a class in $H^1(\Sigma,\Omega^{1,cl})$. These
likewise will make a later appearance.

In perturbation theory, it suffices for the conditions \invi\ to
hold at the level of differential forms.  Nonperturbatively (and
thus beyond the scope of the present paper), these conditions must
hold in integral cohomology.  For a brief elucidation of this (and
an implementation of the Green-Schwarz anomaly cancellation by
which one defines the fermion path integral once conditions like
\invi\ are imposed), see section 2.2 of \ref\egwitten{E. Witten,
``World-Sheet Corrections Via $D$-Instantons,'' JHEP
0002:030,2000, hep-th/9907041.}.

Both of these anomalies are familiar in closely related models.
The $p_1(X)$ anomaly appears equally with $(0,1)$ or $(0,2)$
supersymmetry  and is quite important in the context of the
heterotic string.  The $c_1(\Sigma)c_1(X)$ anomaly appears in
sigma models with  $(2,2)$ supersymmetry twisted to get the
topological $B$-model and is the reason that the $B$-model (except
in genus 1) is only consistent on Calabi-Yau manifolds.  The
$c_1(\Sigma)c_1(X)$ anomaly, in models with $(0,2)$ or $(2,2)$
supersymmetry, is generated by the topological twist \who, while
the $p_1(X)$ anomaly is present in the underlying physical model
with $(0,1)$ or $(0,2)$ supersymmetry, regardless of any
topological twisting.

\bigskip\noindent{\it Other Questions Involving Anomalies}

Finally, we will just briefly hint at a few other questions
involving anomalies that are important for a more complete study
of the model.

One basic issue is to show that $Q$ is conserved at the quantum
level and that $Q^2=0$.

Assuming that $ Q$ is conserved, the fact that $Q^2=0$ follows
from the fact that for a generic hermitian metric on $X$, there
are no locally defined conserved quantities in the model of charge
$q=2$ with respect to the $U(1)$ symmetry $R$ introduced in
section 2.1. It suffices to show this classically (as small
quantum corrections can only destroy conservation laws, not create
them). So it suffices to show that for generic metric there are no
nontrivial local conserved currents of $q=2$. (A trivial conserved
current is $J=*dC$, where $C$  is a local operator of dimension
$(0,0)$; the associated conserved charge vanishes.) Indeed, it is
possible to show that such currents exist if and only if there are
suitable covariantly constant tensors on $X$ beyond the metric
tensor.

This can be contrasted with what happens for the bosonic string
outside the critical dimension.  The BRST operator $Q$ is
conserved, but its square is nonzero and is a multiple of the
conserved quantity $\oint dz\,c \partial^3c$.

To show that $ Q$ is conserved, one approach is to note that
$Q$-invariant Pauli-Villars regulator terms $\int d^2x \{Q, V\}$
are possible (where $V$ is a suitable higher derivative expression
of $R$-charge $-1$, such as $g_{i\bar j}\rho_{\bar
z}^i\partial_{\bar z}\partial^2_z\bar\phi{}^{\bar i}$). Upon
adding such terms, all Feynman diagrams are regularized beyond
one-loop order, so only one-loop anomalies are possible. To show
that there is no one-loop anomaly in $Q$ requires some more direct
argument. One approach is to classify, in terms of the local
differential geometry, the possible anomalies that might appear at
one-loop order, and show that there are none.  (Here one uses the
fact that in general, sigma model perturbation theory is local on
$X$, and the one-loop approximation only involves derivatives of
the metric of $X$ to very low order.  A similar argument can
actually be carried out to all orders, to show that $ Q$ is
conserved without using the fact that a regularization exists
beyond one-loop order.)

The proof of conservation of $Q$ can be carried out in the
presence of an arbitrary metric $f_{ab}$ on $\Sigma$, not just the
flat metric that we have used in writing many formulas. Since the
effective action $\Gamma$ is thus $Q$-invariant for any metric, it
follows that the stress tensor, whose expectation value is
$\langle T_{ab}\rangle=\partial\Gamma/\partial f^{ab}$, is
likewise $Q$-invariant.  The components $T_{\bar z\,\bar z}$ and
$T_{z\bar z}$ of the stress tensor have dimensions $(0,2)$ and
$(1,1)$. On a flat Riemann surface $\Sigma$, the $Q$-cohomology
vanishes for operators of dimension $(n,m)$ with $m\not= 0$, as we
discussed in section 2.1.  Hence, on a flat Riemann surface, we
have $T_{\bar z\,\bar z}=\{Q,\dots\}$ and $T_{z\bar
z}=\{Q,\dots\}$. On a curved Riemann surface $\Sigma$, we have to
allow operators that depend on the Ricci scalar $R$ of $\Sigma$
(which we consider to have dimension $(1,1)$ because of the way it
scales under rescaling of the metric $f$) as well as  its
derivatives, in addition to the usual quantum fields.  In
particular, in this enlarged sense, the $Q$-cohomology of
operators of dimension $(1,1)$ is one-dimensional, being generated
by $R$ itself.  So the general result is \eqn\genres{T_{z\bar
z}={c\over 24\pi}R+\{Q,\dots\},} where $c$ is a constant.  In
particular, though classically $T_{z\bar z}=0$, reflecting
conformal invariance, quantum mechanically there may be an
anomaly.  The anomaly is the sum of a multiple of $R$,
corresponding to the usual conformal anomaly, and a $Q$-trivial
term that, being $Q$-trivial, does not affect correlation
functions of operators in the chiral algebra, that is, operators
annihilated by $Q$.  The $c$-number anomaly can be considered to
affect only the partition function, not the normalized correlation
functions. Combining the statements in this paragraph, correlation
functions of operators in the $ Q$-cohomology are holomorphic and
depend on $\Sigma$ only via its complex structure,  as is familiar
for chiral algebras.

\newsec{Sheaf of Perturbative Observables}

In this section, we analyze the $Q$-cohomology in perturbation
theory.  Nonperturbatively,  and beyond the scope of the present
paper, instanton corrections can change the picture radically.

\subsec{General Considerations}

A local operator is represented by an operator $F$ that in general
is a function of the fields $\phi$, $\bar\phi$, $\rho_{\bar z}$,
$\alpha$, and their derivatives with respect to $z$ and $\bar
z$.\foot{In contrast to section 2.3, here we work locally on a
flat Riemann surface with local parameter $z$, so we need not
include in our operators dependence on the scalar curvature of
$\Sigma$.} However, as we saw in section 2.1, the $Q$-cohomology
vanishes for operators of dimension $(n,m)$ with $m\not= 0$. Since
$\rho_{\bar z}$ and the derivative $\partial_{\bar z}$ both have
$m=1$ (and no ingredient in constructing a local operator has
negative $m$), $Q$-cohomology classes can be constructed from just
$\phi$, $\bar\phi$, $\alpha$, and their derivatives with respect
to $z$. The equation of motion for $\alpha$ is $D_z\alpha=0$, so
we can ignore the $z$-derivatives of $\alpha$.  A $Q$-cohomology
class can thus be represented in general by an operator
\eqn\jurpo{F(\phi,\partial_z\phi,\partial_z^2\phi,\dots;
\bar\phi,\partial_z\bar\phi,\partial_z^2\bar\phi,\dots; \alpha),}
where we have tried to indicate that $F$ might depend on $z$
derivatives of $\phi$ and $\bar\phi$ of arbitrarily high order,
though not on derivatives of $\alpha$. If $F$ has bounded
dimension, it depends only on derivatives up to some finite order
and is polynomial of bounded degree in those. $F$ is also
polynomial in $\alpha$, simply because $\alpha$ is fermionic and
only has finitely many components. However, the dependence of $F$
on $\phi$ and $\bar \phi$ (as opposed to their derivatives) is not
restricted to have any simple form. Recalling the definition of
the $R$-charge in section 2.1, we see that if $F$ is homogeneous
of degree $k$ in $\alpha$, then it has $R$-charge $q=k$.

A general $q=k$ operator
$F(\phi,\partial_z,\phi,\dots;\bar\phi,\partial_z\bar\phi,\dots;\alpha)$
can be interpreted as a $(0,k)$-form on $X$ with values in a
certain holomorphic vector bundle. We will make this explicit for
operators of dimension $(0,0)$ and $(1,0)$, hoping that this will
make the general idea clear. For dimension $(0,0)$, the most
general $q=k$ operator is of the form
$F(\phi,\bar\phi;\alpha)=f_{\bar j_1,\dots,\bar
j_k}(\phi,\bar\phi)\alpha^{\bar j_i}\dots \alpha^{\bar j_k}$;
thus, $F$ may depend on $\phi$ and $\bar\phi$ but not on their
derivatives, and is $k^{th}$ order in $\alpha$. Mapping $
\alpha^{\bar j}$ to $d\bar\phi{}^{\bar j}$, such an operator
corresponds to an ordinary $(0,k)$-form $f_{\bar j_1,\dots,\bar
j_k}(\phi,\bar\phi)d\bar\phi{}^{\bar j_1}\dots d\bar\phi^{\bar
j_k}$ on $X$. For dimension $(1,0)$, there are two cases. A
dimension $(1,0)$ operator
$F(\phi,\partial_z\phi,\bar\phi;\alpha)=f_{i,\bar j_1,\dots,\bar
j_k}(\phi,\bar\phi)\partial_z\phi^i\alpha^{\bar
j_1}\dots\alpha^{\bar j_k}$ that is linear in $\partial_z\phi$ and
does not depend on any other derivatives is a $(0,k)$-form on $X$
with values in $T^*X$ (the holomorphic  cotangent bundle of $X$);
alternatively, it is a $(1,k)$-form on $X$.  Similarly, a
dimension $(1,0)$ operator
$F(\phi,\bar\phi,\partial_z\bar\phi;\alpha)=f^i{}_{\bar
j_1,\dots,\bar j_k}(\phi,\bar\phi)g_{i\bar
s}\partial_z\bar\phi{}^{\bar s}\alpha^{\bar j_i}\dots \alpha^{\bar
j_k}$ that is linear in $\partial_z\bar\phi$ and does not depend
on any other derivatives is a $(0,k)$-form on $X$ with values in
$TX$, the holomorphic tangent bundle of $X$.  In a like fashion,
for any integer $n>0$, the operators of dimension $(n,0)$ and
charge $k$ can be interpreted as $(0,k$)-forms with values in a
certain holomorphic vector bundle over $X$.  This structure
persists in quantum perturbation theory, but there  may be
perturbative corrections to the complex structure of this bundle.

The action of $Q$ on such operators is easy to describe at the
classical level. If we interpret $\alpha^{\bar i}$ as
$d\bar\phi{}^{\bar i}$, then $Q$ acting on a function of $\phi$
and $\bar \phi$  is simply the $\bar\partial $ operator.  This
follows from the transformation laws $\delta\bar\phi{}^{\bar
i}=\alpha^{\bar i},\,\,\delta\phi^i=0$. Classically, the
interpretation of $Q$ as the $\bar\partial$ operator remains valid
when $Q$ acts on a more general operator
$F(\phi,\partial_z,\phi,\dots;\bar\phi,\partial_z\bar\phi,\dots;\alpha)$
that does depend on derivatives of $\phi$ and $\bar\phi$.  The
reason for this is that, because of the equation of motion
$D_z\alpha=0$, one can neglect the action of $Q$ on derivatives
$\partial_z^m\bar\phi$ with $m>0$.  One is thus left classically
only with the action of $Q$ on $\bar\phi$, as opposed to its
derivatives; this is interpreted as the $\bar\partial$ operator.

Perturbatively, there definitely are corrections to the action of
$Q$.  The most famous such correction is associated with the
one-loop beta function.  Classically, the dimension $(2,0)$ part
of the stress tensor is $T_{zz}=g_{i\bar
j}\partial_z\phi^i\partial_z\bar\phi{}^{\bar j}$.  Classically,
$\{Q,T_{zz}\}=0$, but at the one-loop order,
\eqn\nologo{\{Q,T_{zz}\}=\partial_z(R_{i\bar
j}\partial_z\phi^i\alpha^{\bar j}),} where $R_{i\bar j}$ is the
Ricci tensor.  If $X$ is Calabi-Yau (and thus $R_{i\bar
j}=\partial_i\partial_{\bar j}\Lambda$ for some function
$\Lambda(\phi,\bar\phi)$), it is possible to modify $T_{zz}$
(subtracting $\partial_z(\partial_i\Lambda
\partial_z\phi^i)$) so as to be annihilated by $Q$.  But if
$c_1(X)\not=0$, the one-loop correction to $Q$ is essential and a
$Q$-invariant modification of $T_{zz}$ does not exist.  In section
5, we will examine more closely, from a different point of view,
the one-loop correction to the cohomology of $Q$ that is
associated with the beta function.

Gradually, we will obtain a fairly clear picture of the nature of
perturbative quantum corrections to $Q$.  For now, we make a few
simple observations. Let $Q_{cl}=\bar\partial$ denote the
classical approximation to $Q$. Perturbative corrections to $Q$
are local on $X$; they modify the classical formula
\eqn\jiko{Q_{cl}=\bar\partial=\sum_id\bar\phi{}^{\bar
i}{\partial\over\partial\bar\phi{}^{\bar i}}=\sum_i\alpha^{\bar
i}{\partial\over\partial\bar\phi{}^{\bar i}}} by terms that, order
by order in perturbation theory, are differential operators whose
possible degree grows with the order of perturbation theory. This
is so because, more generally, sigma model perturbation theory is
local on $X$ (and to a given order, sigma model perturbation
theory depends on an expansion of fields such as the metric tensor
of $X$ in a Taylor series up to a given order). Instanton
corrections are not at all local on $X$, so they can change the
picture radically.

The locality highly constrains the possible perturbative
modifications of $Q$. Let us try to perturb the classical
expression \jiko\ to a more general operator $ Q=Q_{cl}+\epsilon
Q'+{\cal O}(\epsilon^2)$, where $\epsilon$ is a small parameter
that controls the magnitude of perturbative quantum corrections.
To ensure that $ Q^2=0$, we need $\{Q_{cl},Q'\}=0$; moreover, if
$Q'=\{Q_{cl},\Lambda\}$ for some $\Lambda$, then the deformation
by $Q'$ can be removed by conjugation with
$\exp(-\epsilon\Lambda)$.  So  $Q'$ represents a $Q_{cl}$ or
$\bar\partial$ cohomology class.  Similarly, the same is true of
any essentially new correction to $Q$ (not determined by lower
order terms) that appears at any order in $\epsilon$.   Moreover,
if $Q'$ is to be generated in sigma model perturbation theory, it
must be possible to construct it locally from the fields appearing
in the sigma model action. (This assertion has no analog for
nonperturbative instanton corrections.)  For example, the Ricci
tensor is constructed locally from the metric of $X$, which
appears in the action, and represents an element of the
$\bar\partial$ cohomology group $H^1(X,T^*X)$, so it obeys these
conditions.  Moreover, we saw in section 2 that it is possible to
perturb the action by an element of $H^1(X,\Omega^{2,cl})$; once
such an element appears in the action,  we certainly might then
expect it to appear in a correction to $Q$ -- we will see more
about this later. But these classes are apparently unique as
one-dimensional $\bar\partial$ cohomology classes on $X$ that can
be constructed locally from fields appearing in the action, and it
may be that in some sense they completely determine the
perturbative corrections to $Q$.

\subsec{A Sheaf Of Chiral Algebras}

In general, as we have seen in section 2, the $Q$-cohomology has
the structure of a chiral algebra with holomorphic operator
product expansions.  In this context, the $Q$-cohomology of
dimension zero plays a special role.  If $f(z)$ and $g(z)$ are
local operators of dimension zero representing $Q$-cohomology
classes, then singularities in the operator product $f(z)g(z')$,
by holomorphy, must be proportional to $(z-z')^{-s}h(z')$ for some
positive integer $s$ and operator $h$ of dimension $-s$.  As there
are no operators of negative dimension in sigma model perturbation
theory, no such operator $h$ exists, and hence the operator
product $f(z)g(z')$ is completely nonsingular as $z\to z'$.  It
follows that, for dimension zero, we can naively set $z=z'$ and
multiply $Q$-cohomology classes to get an ordinary  ring (more
precisely, as some of the operators may be fermionic, this is a
$\Bbb{Z}_2$-graded ring with commutators and anticommutators). We
might call this the chiral ring of the theory, as opposed to its
chiral algebra.\foot{The term chiral ring is most commonly used
for a closely related notion in two-dimensional sigma models with
$(2,2)$ supersymmetry, and in related models in other dimensions.}
In perturbation theory, the chiral ring is actually
$\Bbb{Z}$-graded by the $R$-charge; this grading reduces mod two
to the $\Bbb{Z}_2$-grading just mentioned. (Instantons in general
reduce the $\Bbb{Z}$ grading to a $\Bbb{Z}_{2k}$ grading, where
$2k$ is the greatest divisor of $c_1(X)$.)

If we can assume that $Q$ coincides with $Q_{cl}=\bar\partial$,
then this ring is just the graded ring $H^{0,*}(X)$.  This is
certainly an interesting ring, but it may be ``small.'' For
example, it is finite-dimensional if $X$ is compact.

We can do much better if we realize that in perturbation theory,
because the local operators of the sigma model and the fermionic
symmetry $Q$ can be described locally along $X$, it makes sense to
consider operators that are well-defined not throughout $X$, but
only in a given open set $U\subset X$. Concretely, we get such an
operator if we allow the function
$F(\phi,\partial\phi,\dots;\alpha)$ considered earlier to be
defined only for  $\phi\in U$. $Q$-cohomology classes of operators
defined in an open set $U$ have sensible operator product
expansions (in perturbation theory) involving operators that are
also defined in $U$, and they can be restricted in a natural
fashion to smaller open sets and glued together in a natural way
on unions and intersections of open sets. So we get what is known
mathematically as a ``sheaf of chiral algebras,'' associating a
chiral algebra and a chiral ring to every open set $U\subset X$.
We call this sheaf $\hat{\cal A}$. Nonperturbatively, this
structure will break down since, with instanton effects, neither
the local operators (that is, the operators that are local on the
Riemann surface $\Sigma$) nor their operator product expansions
can be defined locally on $X$.

 The operators that are of dimension
zero and of $R$-charge $q=0$ in a given open set $U\subset X$ are
of special interest.  If $Q=Q_{cl}=\bar\partial$, they are simply
the holomorphic functions on $U$, with the obvious commutative
ring structure.  (This ring, however, does not act on arbitrary
sections of the sheaf $\hat {\cal A}$ over $U$, since a dimension
zero operator $f(\phi)$ may have short distance singularities with
a general chiral operator
$F(\phi,\partial\phi,\dots;\bar\phi,\partial\bar\phi,\dots;\alpha)$.
As a result, the sheaf $\hat {\cal A}$ does not have the natural
structure of a ``sheaf of ${\cal O}$-modules.'')

Starting with this observation, we can show in a variety of ways
that there are no perturbative quantum corrections to $Q$ for
dimension zero and charge zero. For the kernel of a differential
operator $Q$ (the solutions $f$ of $Qf=0$) to have the structure
of a sheaf of commutative rings, $Q$ must be a homogeneous first
order differential operator.\foot{Or conjugate to one by
$\exp(\epsilon D)$ for some differential operator $D$.  Such
conjugacy is inessential and would be removed by change of basis
in the space of local operators.} Though we could deform
$Q_{cl}=d\bar\phi{}^{\bar j}\partial/\partial\bar\phi{}^{\bar j}$
by adding a new term $\epsilon d\bar \phi{}^{\bar j}h^i_{\bar
j}\partial/\partial\phi^i)$, where $h$ represents an element of
$H^1(X,TX)$, this just amounts to deforming the complex structure
of $X$.  Both classically and quantum mechanically, we want to
study the model with arbitrary complex structure on $X$, and we
may as well parameterize the quantum theory by the same complex
structure that we use classically.\foot{In general, a family of
classical field theories with appropriate properties leads to a
family of quantum theories depending on the same number of
parameters, but there is no natural pointwise map between the two
families.} So, with the right parameterization of operators and
theories, we can assume that for dimension and $R$-charge zero,
there is no perturbative quantum correction to $Q$.

A more abstract version of this argument is to assert that, since
the kernel of $Q$ for dimension and charge zero gives a sheaf of
commutative rings, we can define a complex manifold $X'$ as the
``spectrum'' of this ring.  If $Q$ is obtained by perturbative
quantum corrections from $Q_{cl}$, $X'$ is a deformation of $X$;
after possibly reparameterizing the family of quantum theories
that depends on the complex structure of $X$, we can assume that
$X'=X$.

We can reach the same conclusion by showing that there is no
locally constructible cohomology class with the right properties
to describe a deformation of $Q$ for operators of dimension and
charge zero. A correction to $Q$ acting on functions or operators
of dimension and charge zero would have leading term
$d\bar\phi^{\bar j}h^{i_1\dots i_s}_{\bar
j}\partial^s/\partial\phi^{i_1}\dots
\partial\phi^{i_s}$, for some $s>0$.  Here $h$ represents an
element of $H^1(X,{\rm Sym}^sTX)$, where ${\rm Sym}^sTX$ is the
$s$-fold symmetric tensor product of $TX$.  No element of
$H^1(X,{\rm Sym}^sTX)$ can be constructed locally, so $Q$ is
undeformed in acting on functions.

Finally, perhaps the most illuminating proof that $Q$ is
undeformed in its action on functions follows from the description
of the sheaf of chiral algebras that we give in section 3.3.

\bigskip\noindent{\it Description By \v{C}ech Cohomology}

We can alternatively describe the perturbative sheaf of
$Q$-cohomology classes by a sort of \v{C}ech cohomology.  This
will bring us to the mathematical point of view on this subject
\cdo. In fact, we will show that the chiral algebra ${\cal A}$ of
the $Q$-cohomology of the sigma model with target space $X$ can be
computed in perturbation theory as the \v{C}ech cohomology of the
sheaf $\hat {\cal A}$ of locally defined chiral (or $Q$-invariant)
operators. The relation between $\bar\partial$ and \v{C}ech
cohomology is of course standard in ordinary differential
geometry, but here we are working in quantum field theory, and $Q$
does not, in general coincide with the $\bar\partial$ operator.
(As we noted above, there are, in general, nontrivial quantum
corrections involving the Ricci tensor, and perhaps others.)
Nevertheless, thinking of $Q$-cohomology as a generalization of
$\bar\partial$ cohomology, it can be related to \v{C}ech
cohomology by following the standard arguments.

Consider an open set $U\subset X$ that is isomorphic to an open
ball in $\Bbb{C}^n$, where $n={\rm dim}_{\Bbb{C}}(X)$.  Any
holomorphic vector bundle $W\to U$ is trivial and the higher
cohomology $H^q(U,W)=0$ for all $q>0$.  Hence, in particular, in
the classical limit, the $Q$-cohomology of the sheaf of local
operators over $U$ vanishes for $q>0$; and since small quantum
corrections can only annihilate cohomology classes, not create
them, it follows perturbatively that the $Q$-cohomology of local
operators over $U$ likewise vanishes in positive degree.

Now consider a good cover of  $X$ by open sets $U_a$.  Then the
$U_a$ and all of their intersections have the property just
described: $\bar\partial$ cohomology and hence $Q$-cohomology
vanishes in positive degree.

Let $F$ be a $Q$-cohomology class of $q=1$.  We can precisely
imitate the usual arguments about $\bar\partial$ cohomology.  When
restricted to $U_a$, $F$ must be trivial, so $F=\{Q,C_a\}$ where
$C_a$ is an operator of $q=0$ that is well-defined in $U_a$. $C_a$
may depend on $a$, although of course $F$ does not.

Now in the intersection $U_a\cap U_b$, we have
$F=\{Q,C_a\}=\{Q,C_b\}$, so $\{Q,C_a-C_b\}=0$. Let
$C_{ab}=C_a-C_b$.  For each $a$ and $b$, $C_{ab}$ is defined in
$U_a\cap U_b$. Clearly, for all $a,b,c$, we have
\eqn\polkj{C_{ab}=-C_{ba},~~C_{ab}+C_{bc}+C_{ca}=0.}

The sheaf $\hat {\cal A}$ of chiral operators has for its local
sections  the $\alpha$-independent local operators
$F(\phi,\partial_z\phi,\dots;\bar\phi,\partial_z\bar \phi,\dots)$
that are annihilated by $Q$. Each $C_{ab}$ is a section of
$\hat{\cal A}$ over the intersection $U_a\cap U_b$. The properties
found in the last paragraph means that it is natural to think of
the collection $C_{ab}$ as defining an element of the first
\v{C}ech cohomology group $H_{{\rm \check{C}ech}}^1(X,\hat{\cal
A})$.

Just as in the usual case of relating $\bar\partial$ and \v{C}ech
cohomology, we can run all this backwards. If we are given a
family $C_{ab}$ of elements of $H^0(U_a\cap U_b,\hat{\cal A})$
obeying \polkj, we proceed as follows.  Let $f_a$ be a partition
of unity subordinate to the open cover of $X$ given by the $U_a$.
(We recall that this means that $f_a$ is nonzero only inside
$U_a$, and $\sum_a f_a=1$.) Let $F_a=\sum_c
[Q,f_c]C_{ac}$.\foot{Some regularization of the operator product
of $[Q,f_c(\phi,\bar\phi)]$ with $C_{ac}$ is needed, for example
by normal ordering.} Then in $U_a\cap U_b$, $F_a=F_b$, since
$F_a-F_b=\sum_c[Q,f_c](C_{ac}-C_{bc})= \sum_c[Q,f_c]C_{ab}=0$ (we
used \polkj\ and the fact that $\sum_c f_c=1$).  So the $F_a$ are
equal to each other and hence to a $q=1$ operator $F$ that obeys
$\{Q,F\}=0$ and is globally defined throughout $X$.

The above argument should seem familiar from section 2.2 (or from
any description of the relation between \v{C}ech and
$\bar\partial$ cohomology). What we have done is simply to copy
the standard argument relating $\bar\partial$ and \v{C}ech
cohomology to show that, for $q=1$, the $Q$-cohomology coincides
with the \v{C}ech cohomology of the sheaf $\hat {\cal A}$. Nothing
is special here about $q=1$, and imitating the standard argument,
we learn that this is true for all $q$.   So the chiral algebra
${\cal A}$ is, as a vector space, $\oplus_qH_{{\rm
\check{C}ech}}^q(X,\hat{\cal A})$. Henceforth we generally  omit
the label ``\v{C}ech'' in denoting the cohomology of $\hat{\cal
A}$.

In effect, the physical description via a Lagrangian and a $Q$
operator gives a $\bar\partial$-like description of a sheaf
$\hat{\cal A}$ of chiral algebras and its cohomology.  In the
mathematical literature, this sheaf is studied from the \v{C}ech
point of view. Here, the field $\alpha$ is omitted and locally one
considers operators constructed only from $\phi$, $\bar\phi$, and
their derivatives.  Cohomology classes of positive $q$ are
constructed as \v{C}ech $q$-cocycles. Instead, in the physical
approach, the sheaf appears in a $\bar\partial$-like language,
using the differential $Q$, and classes of degree $q$ are
represented by operators that are $q^{th}$ order in the field
$\alpha$.

In section 3.3, we express in a physical language a few key points
that are made \cdo\ in the mathematical literature starting from
the \v{C}ech viewpoint.

\subsec{Relation To A Free $\beta\gamma$ System}

To begin, we will give a convenient description of the local
structure of the sheaf $\hat {\cal A}$.  That is, we will describe
in a new way  the $Q$-cohomology of operators that are regular in
a small open set $U\subset X$. We assume that $U$ is isomorphic to
an open ball in $\Bbb{C}^n$.

The hermitian metric on $X$ only enters the action in terms of the
form $\{Q,\dots\}$ and so does not affect the $Q$-cohomology.
Hence,  to describe the local structure, we can pick a hermitian
metric that is flat when restricted to $U$.  The action, in
general, also contains terms derived from an element of
$H^1(X,\Omega^{2,cl}(X))$, as we explained in section 2.2.  These
terms are also $Q$-exact locally, and so can be discarded in
analyzing the local structure in $U$.  We can pick coordinates in
$U$ such that the action derived from the flat hermitian metric in
$U$ is \eqn\olopo{I={1\over 2\pi}\int_\Sigma|d^2z|\sum_{i,\bar
j}\delta_{i,\bar j}\left(\partial_{\bar
z}\phi^i\partial_z\bar\phi^{\bar j} +\rho^i\partial_z\alpha^{\bar
j}\right).}

Now let us describe the $Q$-cohomology classes of operators
regular in $U$.  As explained above, these can be represented by
operators of dimension $(n,0)$ that are independent of $\alpha$.
Such operators in general are of the form
$F(\phi,\partial_z\phi,\dots;\bar\phi,\partial_z\bar\phi,\dots)$.
On this class of operators, $Q$ acts as $\alpha^{\bar
j}\partial/\partial\bar\phi{}^{\bar j}$,  and the condition that
$F$ is annihilated by $Q$ is precisely that, as a function of
$\phi$, $\bar \phi$, and their derivatives, it is independent of
$\bar\phi$ (as opposed to its derivatives), and depends only on
the other variables, namely $\phi$ and the derivatives of $\phi$
and $\bar\phi$.\foot{Once again, we can ignore the action of $Q$
on derivatives of $\bar\phi$ because of the equation of motion
$\partial_z\alpha=0$.}  Thus the $Q$-invariant operators are of
the form
$F(\phi,\partial_z\phi,\dots;\partial_z\bar\phi,\partial_z^2\bar\phi,\dots)$.
Differently put, these operators have a general dependence on the
$z$-derivatives of $\phi$ and $\bar\phi$, but in their dependence
on the center of mass coordinate of the string, they are
holomorphic.

If we set $\beta_i=\delta_{i\bar j}\partial_z\bar\phi{}^{\bar j}$,
which is an operator of dimension $(1,0)$, and $\gamma^i=\phi^i$,
of dimension $(0,0)$, then the $Q$-cohomology of operators regular
in $U$ is represented by arbitrary local functions of $\beta $ and
$\gamma$, of the form
$F(\gamma,\partial_z\gamma,\partial_z^2\gamma,\dots;
\beta,\partial_z\beta,\partial_z^2\beta\dots)$. The operators
$\beta$ and $\gamma$ have the operator products of a standard
$\beta\gamma$ system.  The products $\beta\cdot\beta$ and
$\gamma\cdot\gamma$ are nonsingular, while
\eqn\togo{\beta(z)\gamma(w)=-{1\over z-w}+{\rm regular}.} These
statements  can be deduced from the flat action \olopo\ by
standard methods. We can write down an action for fields $\beta$
and $\gamma$, regarded as elementary fields, which reproduces
these OPE's.  It is simply the standard action of the
$\beta\gamma$ system: \eqn\betagaction{I_{\beta\gamma}={1\over
2\pi}\int|d^2z|\sum_i\beta_i\partial_{\bar z}\gamma^i.} The
equations of motion derived from this action assert that
$\partial_{\bar z}\gamma=\partial_{\bar z}\beta=0$.  So a general
local operator of this system is of the form $\tilde
F(\gamma,\partial_z\gamma,\dots;\beta,\partial_z\beta,\dots)$.
Since the theory constructed from the action $I_{\beta\gamma}$ of
the $\beta\gamma$ system reproduces the appropriate list of
operators and OPE's of the sigma model, it follows that the chiral
algebra of the $Q$-cohomology in a small open set $U$ is the same
as the chiral algebra of the $\beta\gamma$ system, restricted to
the same open set.  (Restriction to $U$ just means that the
operator $F$ or $\tilde F$, in its dependence on the zero mode of
$\gamma=\phi$, is required to be holomorphic in $U$, but not
necessarily throughout $X$ or $\Bbb{C}^n$.)

Does the $\beta\gamma$ system reproduce the $Q$-cohomology
globally, or only in a small open set $U$?  First of all,
classically, the action \betagaction\ makes sense globally if we
interpret the fields $\beta$ and $\gamma$ correctly.  $\gamma$
defines a map $\gamma:\Sigma\to X$, and $\beta$ is a $(1,0)$-form
on $\Sigma$ with values in the pull back $\gamma^*(T^*X)$.  With
this interpretation, \betagaction\ becomes the action of what we
might call a nonlinear $\beta\gamma$ system.  Though nonlinear,
this action can be made linear, locally, by choosing local
coordinates $\gamma^i$ on a small open set $U\subset X$. Although
this sort of nonlinear $\beta\gamma$ system is not widely studied
in physics, there are examples where it has been studied, for
example in covariant quantization of the superstring \ref\berk{N.
Berkovits, ``Super Poincar\'e Covariant Quantization Of The
Superstring,'' JHEP 0004:018,2000, hep-th/0001035.}. In that
application, $X$ is a $\Bbb{C}^*$ bundle over the homogeneous
space $SO(10)/U(5)$.

Granted that the classical action of the $\beta\gamma$ system
makes sense globally, what happens quantum mechanically? The
anomalies that enter in the sigma model also appear in the
nonlinear $\beta\gamma$ system. Expand around a classical solution
of the nonlinear $\beta\gamma$ system, represented by a
holomorphic map $\gamma_0:\Sigma\to X$. Setting
$\gamma=\gamma_0+\gamma'$,  the action, expanded to quadratic
order about this solution,  is $(1/2\pi)( \beta ,\bar D\gamma')$.
Here, the kinetic operator is the $\bar D$ operator on sections of
$ \gamma_0^*(TX)$; it is the complex conjugate of the operator
whose anomalies we encountered in section 2.3.  Complex
conjugation reverses the sign of the anomalies, but here the
fields are bosonic, while in section 2.3, they were fermionic;
this gives a second sign change.  So the nonlinear $\beta\gamma$
system has exactly the same anomalies as the underlying sigma
model. In effect, in going from the sigma model to the nonlinear
$\beta\gamma$ system, we have canceled antiholomorphic bosons and
fermions that do not contribute to the $Q$-cohomology and whose
net contribution to anomalies also vanishes.

On the other hand, the nonlinear $\beta\gamma$ system lacks the
$U(1)$ $R$-charge $q$ of the sigma model.  While locally the
$Q$-cohomology is supported at $q=0$, globally there is
generically cohomology in higher degrees.  How would we use the
nonlinear $\beta\gamma$ system to describe this higher cohomology?
The answer should be clear from section 3.2.  In the $\beta\gamma$
description, we do not have a close analog of $\bar\partial$
cohomology at our disposal, but we can use \v{C}ech cohomology. We
cover $X$ by small open sets $U_a$, and, as explained in section
3.2, we describe the $Q$-cohomology classes of positive degree by
\v{C}ech cocycles. Though this is an unusual procedure (in the
present context) from the point of view of physicists, it has been
taken mathematically as the starting point for the present subject
\cdo. (The subject has also been developed mathematically with
less emphasis on the \v{C}ech point of view \bd.)

Perhaps a more severe problem with the nonlinear $\beta\gamma$
action \betagaction\ is that in this framework, it is difficult to
see all the moduli of the sigma model in the classical action. As
we saw in section 2.2, those moduli are the complex structure of
$X$ and also a class in $H^1(X,\Omega^{2,cl}(X))$.  The complex
structure is built into the classical action \betagaction, but it
does not seem possible to build a class in
$H^1(X,\Omega^{2,cl}(X))$ into the action in this framework.  In
the usual mathematical approach \cdo, the class in
$H^1(X,\Omega^{2,cl}(X))$ is instead incorporated into the
definition of \v{C}ech cocycles, as we explain in section 3.5.

Finally, in a quantum field theory, one wants to do more than
define  $Q$-cohomology classes or a sheaf of chiral algebras.  One
wants to compute correlation functions of operators representing
these cohomology classes as well as, possibly, other local
operators.  For the sigma model, there is a clear procedure to
compute correlation functions, while for the nonlinear
$\beta\gamma$ system there is at first sight no natural procedure,
as there is no sensible way to integrate over the zero mode of
$\gamma$.  The reason for this is clear if we consider again the
charge $q$ of the operators.  In perturbation theory, on a Riemann
surface $\Sigma$ of genus $g$, a correlation function $\langle
{\cal O}_1(z_1)\dots {\cal O}_s(z_s)\rangle$ of operators ${\cal
O}_i$ of charge $q_i$ vanishes unless $\sum_iq_i=n(1-g)$. For
instantons, the formula becomes $\sum_iq_i=n(1-g)+\int_{\Sigma}
\Phi^*(c_1(X))$.  (These formulas comes from the index theorem for
the $U(1)$ current associated with the $R$-charge.  The right hand
side is the dimension of instanton moduli space.) Generically,
therefore, nonzero correlation functions require that the $q_i$ do
not all vanish. As operators of $q_i\not= 0$ cannot be represented
in a standard fashion in the nonlinear $\beta\gamma$ system (but
must be described by  \v{C}ech cocycles), it is clear that, while
in the sigma model one can compute correlation functions via a
standard recipe, to do so in the nonlinear $\beta\gamma$ system
requires translating the usual recipe into the \v{C}ech language.
This would be an unusual procedure, at least for physicists.
(Moreover, to compute such correlation functions at the instanton
level requires understanding instanton corrections to the
$Q$-cohomology, which can radically change the picture and are
beyond the scope of the present paper.)

\subsec{Local Symmetries}

Having understood the local structure of the $Q$-cohomology, we
can attempt to build a global picture by gluing together the local
pieces.

We cover $X$ by small open sets $U_a$.  In each $U_a$, the
$Q$-cohomology can be described by a free $\beta\gamma$ system. We
want to glue these local descriptions together in intersections
$U_a\cap U_b$, so as to describe the $Q$-cohomology in terms of a
sheaf of chiral algebras over the whole manifold $X$.

The gluing must be carried out by an automorphism of the free
$\beta\gamma$ system, so we must understand the symmetries of this
system.  The key properties can be understood by constructing the
Lie algebra $\goth{g}$ of such symmetries.  An element of
$\goth{g}$ is the integral of a dimension one current, modulo
total derivatives. The currents in the $\beta\gamma$ system are as
follows.

First, if $V^i$ is a holomorphic vector field on $X$, we can make
the dimension one current $J_V=-V^i\beta_i$ and the corresponding
conserved charge $K_V=\oint J_V$.  Let $\goth{v}$ be the subspace
of $\goth{g}$ generated by the $K_V$'s.  As shown in \cdo, and as
we will explain momentarily, $\goth{v}$ is {\it not} a Lie
subalgebra of $\goth{g}$, only a linear subspace.

By computing operator products with the elementary fields
$\gamma$, \eqn\hro{J_V(z)\gamma^k(w)\sim {V^k(w)\over z-w},} we
see that $J_V$ generates the infinitesimal diffeomorphism
$\delta\gamma^k=V^k$ of $U$.  Thus, the $J_V$ generate the
holomorphic diffeomorphisms of the target space.

The other conserved currents are as follows.  Let $B=\sum_i
B_i\,d\gamma^i$ be a holomorphic $(1,0)$-form on $X$.  Then we can
make the current $J_B=B_i\partial\gamma^i$, and the conserved
charge $\oint J_B$. However, if $B$ is exact, say
$B_i=\partial_iH$ for some local holomorphic function $H$, then
$\oint J_B=\oint
\partial_iH d\gamma^i=\oint dH=0$. So the conserved charged
constructed from $B$ vanishes if (and only if) $B$ is exact.
Locally, $B$ is exact if and only if $0=\partial
B=\partial_iB_j-\partial_jB_i$. (As we are working in perturbation
theory, it suffices to work locally.) We write $C$ for the
holomorphic $(2,0)$-form $C=\partial B$. It is annihilated by
$\partial$ and so is a local holomorphic section of
$\Omega^{2,cl}$. For every local holomorphic section $C$ of
$\Omega^{2,cl}$, we find a local holomorphic $(1,0)$-form $B$ with
$C=\partial B$ and write $K_C=\oint J_B$. Let us write $\goth{c}$
for the linear span of the $K_C$.

So finally, the symmetry algebra $\goth{g}$ of the $\beta\gamma$
system in a small open set $U$ is, as a linear space, $\goth{g}
=\goth{v}\oplus \goth{c}$.  $\goth{c}$ is trivially a subalgebra,
in fact an abelian one, because the currents $J_B$ derived from
$(1,0)$-forms are constructed only from $\gamma$ (and its
derivatives) and their products have no short distance
singularities.  So $\goth{g}$ is an extension \eqn\exto{0\to
\goth{c}\to \goth{g}\to \goth{v}\to 0.} In fact, \exto\ is an
exact sequence of Lie algebras, since as we will see momentarily,
$[\goth{v},\goth{c}]\subset \goth{c}$.

The action of $\goth{v}$ on $\goth{c}$ can be found from the OPE
\eqn\kilo{V^i\beta_i(z) \cdot B_j\partial\gamma^j(w)=-{1\over
(z-w)^2}V^iB_i(w)-{1\over
z-w}\left(V^i(\partial_iB_k-\partial_kB_i)+\partial_k(V^iB_i)\right)\partial\gamma^k.}
The commutator of $K_V$ with $K_{\partial B}$ is the residue of
the simple pole on the right hand side.  In the numerator, we
recognize that
$V^i(\partial_iB_k-\partial_kB_i)+\partial_k(V^iB_i)$ is the same
as $({\cal L}_V(B))_k$, which represents the Lie derivative of the
vector field $V$ acting on the one-form $B$.  This is what we
might have guessed based on the result \hro\ showing that the
$J_V$ generate diffeomorphisms of $U$.

However, in the commutator of two elements of $\goth{v}$, we get a
surprise \cdo.  Let $V$ and $W$ be two holomorphic vector fields
on $U$. We compute \eqn\bilo{J_V(z)J_W(w)\sim
-{\partial_jV^i\partial_iW^j(w)\over
(z-w)^2}-{(V^i\partial_iW^j-W^i\partial_iV^j)\beta_j\over z-w}
-{(\partial_k\partial_jV^i)(\partial_iW^j\partial\gamma^k)\over
z-w}.} The first term on the right hand side, being a double pole,
does not contribute to the commutator.  The second and third terms
take values in $\goth{v}$ and $\goth{c}$, respectively.  The
second term, which comes from a single contraction of elementary
fields in evaluting the OPE, is the expected result
$J_V(z)J_W(w)\sim J_{[V,W]}/(z-w)$, where
$[V,W]^k=V^i\partial_iW^k-W^i\partial_iV^k$ is the commutator of
the vector fields $V$ and $W$.  We would get the same result by
computing the commutator of $J_V$ and $J_W$ via Poisson brackets
in the classical $\beta\gamma$ theory.  Like all anomalies in
conformal field theory, the third term comes from a multiple
contraction.  This last term means that $\goth{v}$ does not close
upon itself as a Lie algebra -- the commutator of two elements of
$\goth{v}$ is not contained in $\goth{v}$.  So $\goth{g}$ is not
 a semidirect product of $\goth{v}$ with $\goth{c}$, and the
 extension of Lie algebras in \exto\ is nontrivial.

 \subsec{Gluing
The Open Sets Together}

Now take a suitable  collection of small open sets $U_a\subset
\Bbb{C}^n$. We wish to glue them together to make a good cover of
a complex manifold $X$. On each $U_a$, the sheaf $\hat{\cal A}$ of
chiral algebras is defined by a free $\beta\gamma$ system. We want
to glue together these free conformal field theories to get a
globally defined sheaf of chiral algebras.  Two questions arise:
Is there an obstruction to this gluing?  And if we can carry out
the gluing, what are the moduli of the resulting sheaf?

Let us first recall how this is done geometrically. For each
$a,b$, we pick an open set $U_{ab}\subset U_{a}$, and likewise an
open set $U_{ba}\subset U_b$, and a holomorphic diffeomorphism
$f_{ab}$ between them \eqn\ono{f_{ab}:U_{ab}\cong U_{ba}.} We take
$f_{ba}=f_{ab}^{-1}$. We want to identify a point $P\in U_{ab}$
with a point $Q\in U_{ba}$ if $Q=f_{ab}(P)$. This makes sense if
for any $U_a$, $U_b$, and $U_c,$ we have
\eqn\bono{f_{ca}f_{bc}f_{ab}=1} wherever all the maps are defined
(the space in which they are all defined is what we interpret as
the triple intersection $U_{abc}$). This relation says that the
different pieces $U_a$ can be glued together via the holomorphic
maps $f_{ab}$ to make a complex manifold $X$. Complex moduli of
$X$ appear as parameters in the $f_{ab}$.

Now suppose that we have a sheaf of chiral algebras on each $U_a$.
We want to glue them together on overlaps to get a sheaf of chiral
algebras on $X$.  The gluing must be done using a symmetry not of
the complex manifolds $U_a$, but rather using a symmetry of the
conformal field theories.  So for each pair $U_a$ and $U_b$, we
pick a conformal field theory symmetry $\hat f_{ab}$ that maps the
free $\beta\gamma$ system on $U_a$, restricted to $U_{ab}$, to the
free $\beta\gamma$ system on $U_b$, similarly restricted to
$U_{ba}$. We get a global sheaf of chiral algebras if the gluing
is consistent: \eqn\bonob{\hat f_{ca}\hat f_{bc}\hat f_{ab}=1.}

If we do have a consistent gluing, it makes sense to ask what the
target space is.  The reason that this makes sense is that, as in
\exto, there is a Lie algebra homomorphism $\goth{g}\to \goth{v}$
that ``forgets'' the abelian symmetries in $\goth{c}$ and only
``remembers'' how a symmetry acts geometrically, that is as an
element of $\goth{v}$.  Similarly, at the group level, there is a
map from a conformal field theory gluing operator $\hat f_{ab}$ to
the corresponding geometrical symmetry $f_{ab}$.  The relation
\bonob\ for the $\hat f$'s implies a similar relation  \bono\ for
the $f$'s, so every way to glue together the conformal field
theories determines a geometrical gluing of the $U_a$ to make a
complex manifold $X$ that we call the target space of the
conformal field theory.

However, if the $f_{ab}$ are given, the $\hat f_{ab}$ are not
uniquely determined.  We can still pick, for each $U_{ab}$, an
element $C_{ab}\in H^0(U_{ab},\Omega^{2,cl})$, representing an
element of $\goth{c}$.  Then we transform $\hat f_{ab}\to \hat
f'_{ab}= \exp(C_{ab})\hat f_{ab}$. The condition that the gluing
identity \bonob\ is still obeyed is that in each triple
intersection $U_{abc}$ we should have
\eqn\cono{C_{ab}+C_{bc}+C_{ca}=0.} The $C$'s, in other words, must
define an element of the \v{C}ech cohomology group
$H^1(X,\Omega^{2,cl}(X))$.  Passing from $\hat f$ to $\hat f'$
does not change the target space $X$ -- since $C$ is ``forgotten''
when we project from $\hat f_{ab}$ to the geometrical gluing data
$f_{ab}$.  So we get in this way a family of sheaves of chiral
algebras, with the same target space $X$, and admitting an action
of $H^1(X,\Omega^{2,cl}(X))$.  (The last statement simply means
that given such a sheaf and an element $C\in
H^1(X,\Omega^{2,cl}(X))$, one can make a new sheaf by $\hat f\to
\exp(C)\hat f$.)

The simple form of the cocycle condition \cono\ may require some
explanation.  Why can we omit the $f$'s in writing it?  In the
beginning of this section, we recalled how to build up a complex
manifold $X$ by gluing together abstract open sets $U_a$, using
the gluing maps $f_{ab}$.   Once this is done, by the time one
gets to \v{C}ech cohomology, one usually regards the $U_a$ as
subspaces of a common space $X$, and then it is customary to
suppress the $f$'s in writing the condition \cono\ of a \v{C}ech
cocycle.   The $f$'s would return in the formula if we persist in
regarding the $U_a$'s as subsets of  abstract $\Bbb{C}^n$'s.

\bigskip\noindent{\it The Anomaly}

There is also a possibility here, as in section 2.2, for an
anomaly.  In the present context, this will appear as an
obstruction to the gluing.

Suppose we are given a set of gluing data $f_{ab}$ which obeys
\bono.  There is no natural way to ``lift'' the $f_{ab}$ to
conformal field symmetries $\hat f_{ab}$.   Pick any way to do it.
Though the geometrical relation $f_{ca}f_{bc}f_{ab}$ is obeyed,
the analogous lifted relation may not be. In general, we will have
\eqn\hoffo{\hat f_{ca}\hat f_{bc}\hat f_{ab}=\exp(C_{abc})} for
some $C_{abc}\in H^0(U_{abc},\Omega^{2,cl})$. The reason for
\hoffo\ is that, as the left hand side maps to the identity if
projected to the group of geometrical  symmetries, it must be an
element of the abelian group (generated by $\goth{c}$) that acts
trivially on the coordinates $\gamma^i$ of the $U_a$.

The choice of $\hat f_{ab}$ was not unique.  If we transform $\hat
f_{ab}\to \exp(U_{ab})\hat f_{ab}$, we get \eqn\huno{C_{abc}\to
C'_{abc}= C_{abc}+C_{ab}+C_{bc}+C_{ca}.} If it is possible to pick
the $C_{ab}$ to set all $C_{abc}'=0$, then there is no anomaly and
one can obtain a globally defined sheaf of chiral algebras.

In any event, in quadruple overlaps $U_a\cap U_b\cap U_c\cap U_d$,
the $C$'s obey \eqn\tonon{C_{abc}-C_{bcd}+C_{cda} -C_{dab}=0.}
Along with the equivalence relation \huno, this means that the
$C$'s define an element of the sheaf cohomology group $H^2(X,
\Omega^{2,cl}(X))$.

In section 2.2, we obtained from the sigma model an anomaly
measured by $p_1(X)$, which in de Rham cohomology (that is, in
perturbation theory) represents an element of
$H^2(X,\Omega^{2,cl}(X))$.\foot{This assertion was explained in a
footnote in section 2.3.}
 It has been shown \ngog\ that the
obstruction, associated with the $C$'s, to gluing the free
$\beta\gamma$ systems on the $U_a$ into a global sheaf of chiral
algebras is indeed given by $p_1(X)$.  In section 5.2, we
illustrate this in an example.

\bigskip\noindent{\it The Other Anomaly}

In section 2.3, we really had two anomalies, one involving
$p_1(X)$ while the other was proportional to $c_1(\Sigma)c_1(X)$.
We recall that $\Sigma$ is the Riemann surface on which our
quantum field theory is defined, and $X$ is the target space. How
do we see the second anomaly in the present discussion?

So far, we have constructed a sheaf of chiral algebras globally in
$X$, but only locally along $\Sigma$.  When we defined in section
2.1 a chiral algebra on the $Q$-cohomology of a $\sigma$-model,
conformal invariance was not one of the axioms (see the
next-to-last paragraph of section 2.1).  The reason for this is
that, generically, the $\sigma$-model used to construct the chiral
algebra is not invariant under holomorphic reparameterizations of
the Riemann surface $\Sigma$. We noted in section 3.1 that the
holomorphic part of the stress tensor $T_{zz}$ does not correspond
to an element of the $Q$-cohomology unless $c_1(X)=0$. This gives
an obstruction to reparameterization invariance of the
$Q$-cohomology, and thus of the chiral algebra.

With the reformulation by a $\beta\gamma$ system, it may appear
that the problem of lack of conformal invariance has disappeared.
We are now deriving the chiral algebra, locally on $X$, from a
free $\beta\gamma$ system.  The {\it free} $\beta\gamma$ system is
certainly conformally invariant, and can be defined on an
arbitrary global Riemann surface $\Sigma$.

There is no contradiction here.  The anomaly we are looking for is
proportional to $c_1(\Sigma)c_1(X)$, so it vanishes if we work
locally on $X$ (using a free $\beta\gamma$ system) even if we work
globally on $\Sigma$.  It also vanishes if we work locally on
$\Sigma$ even if we work globally on $X$.  The latter is what we
did in our above discussion of conformal field theory gluing
relations such as \bonob. We will only see the $c_1(\Sigma)c_1(X)$
anomaly if we work globally on both $\Sigma$ and $X$.

The global $\beta\gamma$ system \betagaction\ with target space
$X$ is not quite conformally invariant, as we will make most
explicit in section 5.1 when we consider an example in detail. The
problem arises because of normal-ordering problems in defining
quantum operators corresponding to classical expressions
$F(\gamma,\partial_z\gamma,\dots;\beta,\partial_z\beta,\dots)$. In
comparing different methods of normal-ordering a given operator
that classically has dimension $d$, with the different methods
corresponding to different choices of local coordinates on
$\Sigma$ or $X$, we get results that differ by operators of
dimension no greater than $d$. This corresponds to the statement
that in perturbative quantum field theory, the set of all
classical operators of dimension no greater than $d$ (for any
given integer $d$) can be consistently renormalized, without
considering operators of higher dimension.  Thus, working globally
on $\Sigma$ and $X$, the space of quantum operators is
``filtered'' (but possibly not graded) by the dimension.

Suppose that we cover $\Sigma$ with small open sets $P_\tau$ while
covering $X$ with small open sets $U_a$.  On each $P_\tau$, we
define a free $\beta\gamma$ system with target $U_a$.  Now we want
to glue together the $U_a$'s and $P_\tau$'s to get a chiral
algebra, with target $X$, defined on all of $\Sigma$.  By a
``chiral algebra,'' we mean a system of holomorphically varying
local operators on $\Sigma$, with operator product expansions that
have singularities of the usual type only on the diagonal, and
obeying associativity.  (Missing is the usual claim in physical
discussion of chiral algebras that the chiral algebra on $\Sigma$
comes from a universal, conformally invariant chiral algebra that
is universally defined on all Riemann surfaces and has been
specialized to $\Sigma$.)

In effect, we are covering $X\times \Sigma$ with open sets
$W_{a\tau}=U_a\times P_\tau$.  On each such open set, we define a
free $\beta\gamma$ system and hence a chiral algebra, and then on
overlaps we want to glue these together.  The gluing is made using
a combination of holomorphic changes of coordinate on $\Sigma$
(since the free $\beta\gamma$ system is conformally invariant) and
holomorphic changes of coordinate in the target space $X$ (since
the free $\beta\gamma$ system has the geometrical symmetries
$\goth{v}$ generated by vector fields). There is no problem in
finding a gluing map $g_{a\tau,b\nu}$ from $W_{a\tau}$ to
$W_{b\nu}$, but there may be a problem in arranging on triple
overlaps to get suitable relations
$g_{c\sigma,a\tau}g_{a\tau,b\nu}g_{b\nu,c\sigma}=1$. The
obstruction will be a two-dimensional \v{C}ech cohomology class
$H^2(X\times \Sigma,{\cal U})$ for some sheaf ${\cal U}$ that we
must determine.

${\cal U}$ is a sheaf of symmetries of the free $\beta\gamma$
system, since the problem arises from the indeterminacy in the
gluing maps $g_{a\tau,b\nu}$.  As in our discussion of the anomaly
involving $p_1\in H^2(X,\Omega^{2,cl}(X))$, ${\cal U}$ will be a
sheaf of symmetries that act trivially on the coordinates $\gamma$
of the target space $X$.  The reason for this is that, as the
notion of a map $\gamma:\Sigma\to X$ makes sense, there is no
inconsistency in the gluing of $\gamma$.

We earlier identified the abelian group $\goth{c}$ of symmetries
that act trivially on $\gamma$.  However, in that discussion we
implicitly assumed reparameterization invariance on $\Sigma$,
which is valid locally on $\Sigma$ but perhaps not globally. If we
drop reparameterization invariance, we can write more general
holomorphic operator-valued $(1,0)$-forms on $\Sigma$.  We can
take any holomorphic operator ${\cal O}(z)$ of dimension $(n,0)$
for any integer $n$,  and any holomorphic section $f(z)$ of an
appropriate power of the canonical bundle of $\Sigma$ (restricted
to a suitable open set $P_\tau$) and consider the symmetry
generated by $\oint f(z){\cal O}(z)$.  However, because we are
looking for deformations in which the local operators are filtered
by dimension, there is a drastic simplification: we can limit
ourselves to $n\leq 1$. Also, since the ambiguity in gluing comes
from operators that commute with $\gamma$, and so are not visible
geometrically, we need only consider operators constructed  from
$\gamma$ and its derivatives, and not depending on $\beta$; the
constraint $n\leq 1$ means that derivatives enter only via a
linear dependence on $\partial_z\gamma$.

The most general symmetry obeying these conditions is generated by
$\oint J$, with \eqn\hsnn{J=B_i(\gamma,z)d\gamma^i
+E(\gamma,z)dz.} Here $B_i$ and $E$ are the components of a
holomorphic $(1,0)$-form on $X\times \Sigma$, namely
$Y=B_id\gamma^i+E\,dz$.

The  conserved charge  $K_Y=\oint(B_id\gamma^i+E\,dz)$ vanishes
if, and only if, $Y$ is exact, $Y=\partial\Lambda$ for some
zero-form $\Lambda(\gamma,z)$ on $X\times \Sigma$. So this charge
really depends only on the closed holomorphic $(2,0)$-form
$Z=\partial Y$ on $X\times \Sigma$.  $Z$ is a local holomorphic
section of $\Omega^{2,cl}(X\times \Sigma)$. Thus, the we learn
finally that the appropriate sheaf of symmetries that act
trivially on $\gamma$ is isomorphic to
$\Omega^{2,cl}(X\times\Sigma)$.

We therefore must expect that an anomaly takes values in
$H^2(X\times\Sigma,\Omega^{2,cl}(X\times\Sigma))$.  Indeed,
$c_1(X)c_1(\Sigma)$ takes values in this group, as $c_1(X)\in
H^1(X,\Omega^{1,cl}(X))$, $c_1(\Sigma)\in
H^1(\Sigma,\Omega^{1,cl}(X))$, and the wedge product of these
takes values in $H^2(X\times\Sigma,\Omega^{2,cl}(X\times\Sigma))$.
So it is natural for the anomaly seen using Dirac operators in the
sigma model to have an alternative interpretation in terms of the
\v{C}ech cohomology of the sheaf of chiral algebras.  This
description has been developed in the mathematical literature.

Actually, as $\Sigma$ is of complex dimension one, we have
$\Omega^{2,cl}(X\times\Sigma)=\Omega^{2,cl}(X)\otimes {\cal
O}_\Sigma\oplus \Omega^{1,cl}(X)\otimes \Omega^{1,cl}(\Sigma)$.
(Here ${\cal O}_\Sigma$ is the sheaf of holomorphic functions on
$\Sigma$.)  Hence for compact $\Sigma$, \eqn\lollypop{H^2(X\times
\Sigma,\Omega^{2,cl}(X\times
\Sigma))=H^2(X,\Omega^{2,cl}(X))\oplus
H^1(X,\Omega^{1,cl}(X))\otimes
H^1(\Sigma,\Omega^{1,cl}(\Sigma))\oplus \dots.} (We have used the
fact that $H^0(\Sigma,{\cal O})\cong\Bbb{C}$.)  The two anomalies
$p_1(X)/2$ and $c_1(X)c_1(\Sigma)$ take values in the two summands
on the right hand side of \lollypop.

Similarly, if the anomaly vanishes and there exists a global sheaf
of chiral algebras on $\Sigma$, with target space $X$, then moduli
of this sheaf are parameterized (apart from the obvious
geometrical moduli) by
$H^1(X\times\Sigma,\Omega^{2,cl}(X\times\Sigma))$.

\newsec{$(0,2)$ Supersymmetry}

\subsec{Construction Of Models}

The reason the structure explored in this paper is relevant to
physics is that it arises in sigma models with $(0,2)$
supersymmetry. These are unitary, or physically sensible, quantum
field theories, and they have applications for compactification of
the heterotic string. $(0,2)$ supersymmetry requires considerably
stronger conditions than we assumed in sections 2 and 3, where we
considered a general hermitian metric on a complex manifold $X$
and constructed a model with a single fermionic symmetry $Q$
obeying $Q^2=0$.\foot{By contrast, weaker conditions are needed
for $(0,1)$ supersymmetry.  In this case, no complex structure is
required on the target space, and the curvature of the $B$-field
is an arbitrary closed three-form.  This model  has one conserved
supercharge, superficially like the model considered in sections 2
and 3, but the supercharge is hermitian and its square is not
zero.} In $(0,2)$ supersymmetry, one has a pair of fermionic
symmetries, which are hermitian adjoints of one another and
roughly are loop space analogs of the $\bar\partial$ and
$\bar\partial^\dagger$ operators of a finite-dimensional complex
manifold.  Thus, mathematically, if one wants an analog of Hodge
theory for CDO's, $(0,2)$ supersymmetry is natural.

\nref\hull{C. Hull and E. Witten, ``Supersymetric Sigma Models And
The Heterotic String,''
Phys. Lett. {\bf B160} (1985) 398.}%
\nref\ds{M. Dine and N. Seiberg, ``$(2,0)$ Superspace,'' Phys.
Lett. {\bf B180} (1986) 364.}%
In general, a hermitian metric $g$ on a complex manifold $X$
determines an associated $(1,1)$-form $\omega$.  On a Kahler
manifold, $\omega$ is closed, while for $(0,2)$ supersymmetry, it
must obey $\bar\partial\partial\omega=0$.  The present section
contains no novelty; we merely summarize some familiar results
\refs{\hull,\ds} about $(0,2)$ supersymmetry in order to make
clear how the subject of the present paper is related to physics.

To construct a model with $(0,2)$ supersymmetry, we enlarge the
worldsheet $\Sigma$ to a supermanifold $\hat \Sigma$ with bosonic
coordinates $z,\bar z$ and fermionic coordinates $\theta^+$,
$\bar\theta{}^+$.\foot{The reason for the superscripts $+$ is that
$\theta^+$, $\bar\theta{}^+$ transform as sections of one of the
spin bundles of $\Sigma$, say the one of positive chirality. In a
model with $(2,2)$ supersymmetry, one would have additional
fermionic coordinates $\theta^-$, $\bar \theta{}^-$ of the
opposite type.} The supersymmetries act geometrically:
\eqn\ivo{\eqalign{\bar Q_+& =
{\partial\over\partial\bar\theta{}^+}-i\theta^+{\partial\over\partial\bar
z}\cr  Q_+& =
{\partial\over\partial\theta^+}-i\bar\theta{}^+{\partial\over\partial
\bar z}.\cr}} Thus, $Q_+^2=\bar Q_+^2=0$ and $\{Q_+,\bar
Q_+\}=-2i\partial/\partial\bar z$. To construct Lagrangians
invariant under $Q_+$ and $\bar Q_+$, we use the fact that these
operators commute with the supersymmetric derivatives
\eqn\ivo{\eqalign{\bar D_+& =
{\partial\over\partial\bar\theta{}^+}+i\theta^+{\partial\over\partial\bar
z}\cr  D_+& =
{\partial\over\partial\theta^+}+i\bar\theta{}^+{\partial\over\partial
\bar z},\cr}} as well as with $\partial_z$ and $\partial_{\bar
z}$. Moreover, the measure $|d^2z|d\theta d\bar\theta$ is
supersymmetric, that is, invariant under $Q_+$ and $\bar Q_+$. So
any action constructed using only the supersymmetric derivatives
and measure will be supersymmetric.

We will constuct a supersymmetric model of maps
$\Phi:\hat\Sigma\to X$. We consider maps that are required to be
``chiral.''  This means that if $w$ is any local holomorphic
function on $X$, then $W=\Phi^*(w)$ obeys \eqn\unon{\bar
D_+W=D_+\bar W = 0.} To write formulas, one usually picks local
complex coordinates $\phi^ i$ on $X$, and describes the theory via
``chiral superfields'' $\Phi^i=\Phi^*(\phi^i)$, which obey $\bar
D_+\Phi^i=D_+\bar\Phi{}^{\bar i}=0$, and so admit expansions
\eqn\tomorrow{\eqalign{\Phi^i& = \phi^i+\sqrt 2\theta^+
\psi_+^i-i\bar\theta{}^+\theta^+\partial_{\bar z}\phi^i \cr
\bar\Phi{}^{\bar i}& = \bar\phi{}^{\bar i}-\sqrt 2\,\bar\theta{}^+
\bar\psi_+{}^{\bar i}+i\bar\theta{}^+\theta^+\partial_{\bar
z}\bar\phi{}^{\bar i}. \cr}} (The factors of $\sqrt 2$ are
conventional.)

By acting with $Q_+$ and $\bar Q_+$, defined as in \ivo, we can
determine how the fields transform under supersymmetry.  In
particular, $\bar Q_+$ generates the transformation
\eqn\ino{\eqalign{\delta\phi^i & = 0\cr
                  \delta\bar\phi{}^{\bar i}& = -\sqrt
                  2\,\bar\psi_+{}^{\bar i}\cr
                   \delta\psi_+^i&= -i\sqrt 2 \partial_{\bar
                   z}\phi^i\cr
                   \delta\bar\psi_+^{\bar i}&=0.\cr}}
If we set $\alpha^{\bar i}=-\sqrt 2\, \bar\psi_+{}^{\bar i}$,
$\rho^i=-i\psi_+^i/\sqrt 2$, then these transformations coincide
with the ones we started with in eqn. \nurgo.  So  $(0,2)$
symmetry is a specialization of the structure studied in sections
2 and 3, with $Q$ corresponding to $\bar Q_+$.  In the
specialization to $(0,2)$ supersymmetry, there is  also a second
supersymmetry $ Q_+$ that is hermitian adjoint to $\bar Q_+$.
Here, $\bar Q_+$ and $Q_+$ are somewhat analogous to
$\bar\partial$ and $\bar\partial^\dagger$ on an ordinary complex
manifold.  The symmetry generated by $ Q_+$ is in fact, in
components, \eqn\gino{\eqalign{\delta\phi^i&=\sqrt 2\psi_+^i\cr
             \delta\bar\phi^{\bar i}&=0\cr
                 \delta\psi_+^i&=0\cr
                 \delta\bar\psi_+^{\bar i}&=i\sqrt 2\partial_z\bar\phi^{\bar i}.\cr}}

A Lagrangian is constructed locally by introducing a $(1,0)$-form
$K=K_id\phi^i$, with complex conjugate $\bar K=\bar K_{\bar
i}d\bar\phi^{\bar i}$, and setting
\eqn\muto{I=\int|d^2z|d\bar\theta{}^+d\theta{}^+\left(-{i\over
2}K_i(\Phi,\bar\Phi)\partial_z\Phi^i+{i\over 2}\bar K_{\bar
i}(\Phi,\bar\Phi)\partial_z\bar\Phi^{\bar i}\right).} This is
regarded as a local expression for the action -- so in
manipulating it we are free to integrate by parts and discard
exact forms.  A global description will be clear shortly. The
reason that \muto\ is only a local expression for the action is
that  there are transformations of $K$ that only change the action
density by an exact form.   Hence, describing the action in terms
of $K$ is analogous to describing a Kahler manifold in terms of a
``Kahler potential,'' which is a locally-defined zero-form $t$ in
terms of which the Kahler form can locally be written as
$\omega=-i\partial\bar\partial t$.

The most obvious transformations of $K$ that leave the action
 fixed are \eqn\puto{K\to K+\partial \Lambda,~~ \bar K\to \bar
K-\bar\partial\Lambda,} for any imaginary zero-form $\Lambda$.
Under this change in $K$, the action density changes by the total
derivative $\partial_{ z}\Lambda$, which integrates to zero. Less
obvious (but explained presently) is that the action is also
invariant under $K\to K+K'$, $\bar K\to\bar K+\bar K{}'$, where
$K'$ is a holomorphic differential.

The basic object invariant under the transformations just
described and hence globally defined is the hermitian metric
$ds^2=g_{i\bar j}d\phi^id\bar\phi^{\bar j}$, where
\eqn\nonn{g_{i\bar j}=\partial_{\bar j}K_i+\partial_i\bar K_{\bar
j}.} Associated to this metric is the $(1,1)$-form
\eqn\turmo{\omega={i\over 2}\left(\bar\partial K-\partial\bar
K\right).} In contrast to a Kahler manifold, whose Kahler form
obeys $\partial\omega=\bar\partial \omega=0$, \turmo\ implies the
weaker condition \eqn\gorto{\bar\partial\partial\omega=0,} which
therefore characterizes $(0,2)$ supersymmetry. (This condition
might be compared with the condition defining a Guaduchon metric,
which in complex dimension $n$ is
$\partial\bar\partial(\omega^{n-1})=0$. The two conditions
coincide for $n=2$.)

By virtue of \gorto, the $(2,1)$-form ${\cal H}=2i\partial\omega$
obeys $\partial {\cal H}=\bar\partial{\cal  H}=0$.  It hence can
be interpreted as a class in $H^1(X,\Omega^{2, cl})$.  As we will
see, it plays the role of the class called ${\cal H}$ in section
2.2, by which the sheaf of CDO's of a general complex manifold can
be deformed.  Just like the hermitian metric of $X$, ${\cal H}$ is
severely constrained by $(0,2)$ supersymmetry. While in the
general analysis of section 2.2, ${\cal H}$ can be an arbitrary
element of $H^1(X,\Omega^{2,cl})$, if we wish to specialize to
$(0,2)$ supersymmetry, ${\cal H}$ must be of type $(2,1)$ and
expressible as $2i\partial \omega$ in terms of a positive
$(1,1)$-form $\omega$.

To justify these statements, and further explain how $(0,2)$
supersymmetry relates to the more general structure explored in
sections 2 and 3, we want to express \muto\ in the form $\int
|d^2z|\{Q,V\}$, introduced in section 2.1, and also convert it to
an ordinary Lagrangian, expressed just as an ordinary integral
over $z$ and $\bar z$.  The most straightforward way to do this is
to simply perform the integral over $\theta$ and $\bar\theta$. A
convenient shortcut is to note that for any $X$, we can make the
 replacement \eqn\himvo{\eqalign{\int
|d^2z|d\bar\theta{}^+d\theta{}^+X =\int |d^2z|\left.{\partial^2
\over
\partial\bar\theta{}^+\partial\theta{}^+}X\right|_{\bar\theta{}^+=\theta{}^+=0}
&=\int |d^2z|~\biggl.\bar
D_+D_+X\biggr|_{\theta^+=\bar\theta{}^+=0}\cr =-\int
|d^2z|~\biggl. D_+\bar
D_+X\biggr|_{\theta^+=\bar\theta{}^+=0}.\cr}} The basis for the
first step is that, for a fermionic variable $\theta$, $\int
d\theta \,X=(\partial X/\partial \theta)|_{\theta=0}$.  The basis
for the second step is that the $D$'s differ from the
$\partial/\partial\theta$'s by $\partial_{\bar z}$ terms, which
vanish upon integration by parts. The basis for the third step is
that $\{D_+,\bar D_+\}=2i\partial_{\bar z}$, which again vanishes
upon integration by parts.  We can, for example, now see that
$\int |d^2z|d\theta^+d\bar\theta{}^+\,X$ is invariant to $X\to
X+K'_i(\Phi)\partial_z\Phi^i$, or to $X\to X-\bar K'_{\bar
i}\partial_z\bar\Phi^{\bar i}$, if $K'$ is a holomorphic one-form.
For example, we have $\bar D_+(K'_i(\Phi)\partial_z\Phi^i)=0$, as
$\bar D_+\Phi=0$ and $K'$ is holomorphic, and hence $\int
|d^2z|D_+\bar D_+\left(K'_i(\Phi)\partial_z\Phi^i\right)=0$.
Similarly, $D_+(\bar K'_{\bar i}\partial_z\bar\Phi^{\bar i})=0$,
by virtue of which $\int |d^2z|\bar D_+D_+(\bar K'_{\bar
i}\partial_z\bar\Phi^{\bar i})=0$.

To evaluate the action,  a slight variant of \himvo\ is more
useful. We write the action as \eqn\miox{\int |d^2z|\left.\{\bar
Q_+,[ D_+,X]\}\right|_{\theta=\bar\theta=0}.} This is valid
because, again, $\bar Q_+$ differs from $\bar D_+$ and
$\partial/\partial\bar\theta{}^+$ by a total derivative.   In
section 2.1, we wrote the action as $\int |d^2z|\{Q,V\}$.  Since
we are identifying $\bar Q_+$ with $Q$, we see that $V= D_+X=-i
D_+(K_i(\Phi,\bar\Phi)\partial_z\bar\Phi^i-\bar K_{\bar
i}(\Phi,\bar\Phi)\partial_z\bar\Phi{}^{\bar i})/2$. To evaluate
this, we note that, as $D_+\bar\Phi=0$, we have
$D_+(K_i\partial_z\Phi-\bar K_{\bar i}\partial_z\bar\Phi^{\bar i})
=K_{i,j}D_+\Phi^j\partial_z\Phi^i+K_i\partial_zD_+\Phi^i-\bar
K_{\bar i,i}D_+\Phi^i\partial_z\bar\Phi^{\bar i}$.  After
subtracting the total derivative $\partial_z(K_iD_+\Phi^i)$, which
will not contribute to the action, we get $2iV=-(K_{i,\bar j}+\bar
K_{\bar j,i})\partial_z\bar\Phi^{\bar
j}D_+\Phi^i+(K_{i,j}-K_{j,i})D_+\Phi^j\partial_z\Phi^i$. To set
$\bar\theta{}^+=\theta^+=0$, we just set $\Phi^i=\phi^i$, $\bar
\Phi{}^{\bar i}=\bar\phi^{\bar i}$, and $D_+\Phi^j=\sqrt
2\psi_+^j=-2i\rho^j$, and let $\bar Q_+$ act as in \ino.  So
$V=-\left((\partial_{\bar i}K_i+\partial_i\bar K_{\bar
i})\rho^i\partial_z\bar\phi{}^{\bar i}
-(K_{i,j}-K_{j,i})\rho^j\partial_z\phi^i\right)$.  It is now
straightforward to read off the hermitian metric $g_{i\bar i}$
used in section 2.1 to construct the basic Lagrangian, as well as
the field called $T$ in section 2.2.  We have $g_{i\bar
i}=\partial_i\bar K_{\bar i}+\partial_{\bar i}K_i$, as claimed
above, and $T_{ij}=\partial_iK_j-\partial_jK_i$.  From the last
statement, it follows that the curvature of the two-form field $T$
is ${\cal H}=dT=\bar\partial\partial K=2i\partial\omega$, as
asserted above.

We should note, however, that in physics one defines the two-form
gauge field $B$ and the associated field curvature $H=dB$ a little
differently. One defines $H={\rm Re}({\cal H})$, so that locally
$H=dB$ with $B=-(1/2)(\bar\partial K+\partial \bar K)$.    The
imaginary part of ${\cal H}$ can, in the case of $(0,2)$
supersymmetry, be written as ${\rm Im}({\cal H})=d\omega$, where
$\omega$ is globally defined, so ${\rm Im}({\cal H})$ is
cohomologous to zero.  Thus, the interesting global information
can equally well be described by $H$ or ${\cal H}$.

Finally, the complete action can be written explicitly
  \eqn\ipo{\eqalign{I=\int |d^2z|&\left(K_{i,\bar
j}\partial_{\bar z}\bar\phi{}^{\bar j}\partial_z\phi^i +K_{\bar
i,j}
\partial_{\bar z}\phi^j\partial_z\bar\phi{}^{\bar i}\right.\cr
&\left. -i\left(K_{i\bar j}\bar\psi_+{}^{\bar
j}\partial_z\psi^i+K_{\bar
i,j}\psi_+^j\partial_z\bar\psi_+{}^{\bar i}\right)
+i\left(K_{i,k\bar j}\psi_+^k\bar\psi_+{}^{\bar
j}\partial_z\phi^i-K_{\bar i,j\bar l}\psi_+^j\bar\psi_+^{\bar
l}\partial_z\bar \phi{}^{\bar i} \right)\right).\cr}}

\subsec{An Example}

\nref\spindel{Ph. Spindel, A. Sevrin, W. Troost, and A. Van
Proeyen, ``Complex Structures on Parallelized Group Manifolds And
Supersymmetric Sigma Models,'' Phys. Lett. {\bf B206} (1988) 71,
``Extended Supersymmetric Sigma Models On Group Manifolds, I. The
Complex Structures,'' Nucl. Phys. {\bf B308}
(1988) 662.}%
\nref\rocek{M. Rocek, K. Schoutens, and A. Sevrin, ``Off-Shell WZW
Models In Extended Superspace,'' Phys. Lett. {\bf B264} (1991)
303.}%

A very simple example of an essentially non-Kahler complex
manifold that is the target space of a $(0,2)$ model is $X={\bf
S}^1\times {\bf S}^3$, originally considered in this context in
\refs{\spindel,\rocek}. This example is quite elementary
geometrically but of considerable interest in conformal field
theory.

The complex structure of $X$ can be constructed as follows. By
composing the projection onto the second factor $X\to {\bf S}^3$
with the Hopf fibration $\pi:{\bf S}^3\to {\bf S}^2\cong{\Bbb{
CP}}^1$, whose fibers are copies of ${\bf S}^1$, $X$ can be
fibered over ${\Bbb{CP}}^1$ with fibers $E={\bf S}^1\times {\bf
S}^1$. Giving $E$ the structure of a complex Riemann surface of
genus one, $X$ becomes a complex manifold.

Alternatively, $X$ can be constructed as $\Bbb{C}^2/\Bbb{Z}$,
where $\Bbb{Z}$ acts on coordinates $z^i,$ $i=1,2$ of $\Bbb{C}^2$
by $z^i\to \lambda^nz^i$, with $\lambda$ a nonzero complex number
of modulus less than 1.  The choice of $\lambda$ determines the
complex structure of $E$ in the other description. The two
descriptions are related by simply regarding the $z^ i$ as
homogeneous coordinates of $\Bbb{CP}^1$.

A hermitian form $\omega$ on $X$ that obeys
$\partial\bar\partial\omega=0$ (and corresponds to real $\lambda$)
can be obtained as follows. We will construct $\omega$ to be
invariant under $U(1)\times U(2)$, where $U(1)$ acts by rotation
of ${\bf S}^1$, leaving fixed a one-form $dt$, and $U(2)$ acts on
${\bf S}^3$ commuting with the Hopf fibration. $U(2)$ induces on
the base ${\bf S}^2$ of the Hopf fibration a rotation symmetry
group $SO(3)$.  Let $\omega_0$ be an $SO(3)$-invariant form on
${\bf S}^2$ that integrates to 1.  When pulled back to ${\bf
S}^3$, $\omega_0$ is topologically trivial, and in fact
$\pi^*(\omega_0)=d\zeta$ for a unique $U(2)$-invariant one-form
$\zeta$ (which integrates to 1 on each fiber of the Hopf fibration
and is an ``angular form'' for this fibration). Finally, we let
\eqn\uninno{\omega= dt\wedge \zeta +\pi^*(\omega_0).} To prove
that $\partial\bar\partial\omega=0$, one approach is to note that
this is $d((\bar\partial-\partial)\omega/2)$, and so is
cohomologically trivial and integrates to zero on the
four-manifold ${\bf S}^1\times {\bf S}^3$.  Since it is also
$U(1)\times U(2)$-invariant, it can only integrate to zero if it
vanishes pointwise.  The form $\omega$ is of type $(1,1)$ for the
complex structure on ${\bf S}^1\times {\bf S}^3$, and the
associated hermitian metric is just a ``round'' metric on ${\bf
S}^1\times {\bf S}^3$, which in fact has the full $SO(4)$ rotation
symmetry of ${\bf S}^3$, and not just the $U(2)$ symmetry of the
complex structure.  One way to describe the complex structure of
${\bf S}^1\times {\bf S}^3$ is to say that the forms of type
$(1,0)$ on ${\bf S}^1\times {\bf S}^3$ are generated by
$dt+i\zeta$ and pullbacks of $(1,0)$-forms on $\Bbb{CP}^1$.

If we simply ask for a $U(1)\times SO(4)$-invariant metric on
${\bf S}^1\times {\bf S}^3$, we note at once that such metrics are
determined by two positive numbers, the radii of the two factors.
How do these two parameters enter in the present construction? The
ratio of radii of ${\bf S}^1$ and ${\bf S}^3$ is determined by the
choice of $dt$ (we specified it to be $U(1)$-invariant, but did
not fix the value of $w=\int_{{\bf S}^1}dt$).  The choice of $dt$
is also correlated with the choice of complex structure, since
$\omega$ must be of type $(1,1)$.  Hence, when the complex
structure of ${\bf S}^1\times {\bf S}^3$ is chosen, the ratio of
radii is fixed.  On the other hand, one can rescale the ${\bf
S}^1$ and ${\bf S}^3$ radii by a common positive constant by
multiplying the action \ipo\ by this constant.  So the complex
structure determines the ratio of radii, and leaves one overall
free parameter.

Now let us compute $H={\rm Re}\,{\cal H}$, the curvature of the
$B$-field. This is most conveniently done using the fact that
$H=i(\partial-\bar\partial)\omega$.  First of all,
$\partial\pi^*(\omega_0)=\bar\partial\pi^*(\omega_0)=0$, so
actually $H=-i(\partial-\bar\partial)(\zeta\wedge dt)$.  To
compute $(\partial-\bar\partial)(\zeta\wedge dt)$, we note that
$\zeta\wedge dt$ is a $(1,1)$-form, so that $d(\zeta\wedge
dt)=\upsilon+\bar\upsilon$, where $\upsilon$ is a $(2,1)$-form and
$\bar\upsilon$ is a $(1,2)$-form, and finally
$(\partial-\bar\partial)( \zeta\wedge dt)=\upsilon-\bar\upsilon$.
But explicitly, $d(\zeta\wedge dt)=dt\wedge\pi^*(\omega_0)$.  Here
$\pi^*( \omega_0)$ is of type $(1,1)$, while $dt+i\zeta$ is of
type $(1,0)$ and $dt-i\zeta$ is of type $(0,1)$.  So
$\upsilon=(1/2)(dt+i\zeta)\pi^*(\omega_0)$,
$\bar\upsilon=(1/2)(dt-i\zeta)\pi^*(\omega_0)$, and at last
$H=-i(\upsilon-\bar\upsilon)=\zeta\wedge\pi^*(\omega_0)$.

It follows that  $\int_{{\bf S}^3}H=1$ and in particular $H$ is
topologically non-trivial.  Therefore, to obtain a quantum theory,
we cannot use an arbitrary action of the form \ipo.   We must
multiply this action by a constant chosen so that $\int_{{\bf
S}^3}H=2\pi k$ for some integer $k$, which must be positive so
that the hermitian metric of ${\bf S}^1\times {\bf S}^3$ is
positive.

In fact, $H\sim\zeta\wedge\pi^*(\omega_0)$ is invariant not just
under the $U(2)$ symmetry of the Hopf fibration but under the full
$SO(4)$ rotation symmetry of ${\bf S}^3$.  The hermitian metric
obtained in this construction is likewise $SO(4)$-invariant, as we
noted above.  So the full sigma model has this symmetry (and as a
result \rocek\ has $(0,4)$ supersymmetry, not just the $(0,2)$
supersymmetry that was built into our construction of it).

In fact, as explained in \refs{\spindel,\rocek}, the $U(1)\times
SO(4)$-invariant supersymmetric sigma model of ${\bf S}^1\times{
\bf S}^3$ is simply a product of a WZW model of  the group
$SU(2)\cong {\bf S}^3$ with a free field theory. (The latter is
the product of a free model of ${\bf S}^1$ times a free fermion
system, which arises because the fermions $\psi$ and $\bar\psi$ in
this particular example become free when expressed in a
left-invariant frame on ${\bf S}^3$.) The level of the WZW model
is $k$.  In particular, the parameter $k$, which from the point of
view of the perturbative theory of CDO's is a complex parameter
associated with $H^1({\bf S}^1\times {\bf S}^3,\Omega^{2,cl})\cong
\Bbb{C}$,\foot{By using the fibration of ${\bf S}^1\times {\bf
S}^3$ over $\Bbb{CP}^1$, one can prove that $H^1({\bf S}^1\times
{\bf S}^3,\Omega^{2,cl})$ is one-dimensional with ${\cal
H}=-2i\upsilon \wedge \pi^*(\omega_0)$ as a generator.} must
actually be an integer in order for the model to be well-defined
nonperturbatively.

We will return to this example in section 5.4.

\newsec{Examples Of Sheaves Of CDO's}

In this section,  we analyze some examples of sheaves of CDO's,
aiming mainly to   illustrate the slightly abstract discussion of
section 3.

\subsec{CDO's of $\Bbb{CP}^1$}

For our first example, following section 5.6 of \cdo, we take
$X=\Bbb{CP}^1$.  We work locally on the worldsheet $\Sigma$,
choosing a local complex parameter $z$ and using it, as explained
below, for normal-ordering.

  Of course, $\Bbb{CP}^1$ can be regarded as the
complex $\gamma$-plane plus a point at infinity.  We can usefully
cover it by two open sets, $U_1$ and $U_2$, where $U_1$ is the
complex $\gamma$-plane, and $U_2$ is the complex $\tilde
\gamma$-plane, where $\tilde\gamma=1/\gamma$.

In $U_1$, since it is isomorphic to $\Bbb{C}$, the sheaf of chiral
 operators or CDO's can be described by a single free $\beta\gamma$ system:
\eqn\mino{I={1\over 2\pi}\int|d^2z|\beta\bar\partial\gamma.} Here
$\beta$ and $\gamma$ are fields of dimension $(1,0)$ and $(0,0)$
with the familiar free-field OPE's; there are no singularities in
the operator products $\beta(z)\cdot\beta(z')$ or
$\gamma(z)\cdot\gamma(z')$, while \eqn\ilfo{\gamma(z)\beta(z')\sim
{1\over z-z'}.}

Similarly, the theory in $U_2$ is described by a free
$\tilde\beta\tilde\gamma$ system, with the same dimensions and
action \eqn\mino{I={1\over 2\pi}\int|d^2z|\tilde
\beta\bar\partial\tilde\gamma,} and the same OPE's.

To completely describe the sheaf of chiral operator, we must
explain the gluing between $\beta,\gamma$ and
$\tilde\beta,\tilde\gamma$ in the overlap region $U_1\cap U_2$.
There is nothing we can do to modify the classical relation
\eqn\zino{\tilde\gamma={1\over\gamma}} in any essential way.  The
reason is that, since $\tilde\gamma$ is supposed to have dimension
zero, the right hand side of \zino\ must be a function of $\gamma$
only (and cannot depend on $\beta$ or the derivatives of $\gamma$
or $\beta$).  Hence, \zino\ is a classical gluing relation that
(given that it is consistent) makes an ordinary complex manifold
obtained by gluing together the $U_i$.  As $\Bbb{CP}^1$ has no
complex moduli, any modification of \zino\ would give back an
equivalent result.\foot{Here we can recall an observation from
section 3.2. Even if we consider instead of $\Bbb{CP}^1$ a complex
manifold $X$ that {\it does} have complex moduli, nothing is
gained by assuming quantum corrections to the classical gluing
laws for the dimension zero fields.  Those gluing laws build up a
complex manifold $X'$ that we can call the target of the quantum
theory, and we may as well parameterize the space of quantum
theories in such a way that $X'$ is isomorphic to the underlying
classical manifold $X$.}

Since $\beta$ has dimension 1, one might expect the appropriate
gluing law for $\beta$ to be $\tilde\beta=\beta'$ where
$\beta'=(\partial\gamma/\partial\tilde\gamma)\beta=-\gamma^2\beta$.
This formula is a little ambiguous, because of the existence of a
short distance singularity in the $\gamma-\beta$ operator product.
We resolve such ambiguities, for any differential polynomial in
$\beta$ and $\gamma$, by normal ordering. So in this case,
$\gamma^2\beta$ is a shorthand for
\eqn\ino{\gamma^2\beta(z)=\lim_{z'\to
z}\left(\gamma^2(z')\beta(z)-{2\over z'-z}\gamma(z')\right),}
which to be more precise is also sometimes denoted as
$:\gamma^2\beta:(z)$. This normal ordering definition of a local
operator gives results that depend on the choice of local
parameter $z$ (though the space of all operators does not depend
on this choice).  For this reason, as we discussed at several
junctures in sections 2 and 3, once we construct a chiral sheaf,
its invariance under reparameterizations of $\Sigma$ is not
guaranteed.

However, now that we have specified exactly what it means, the
gluing formula $\tilde\beta=\beta'$ is not right, as we find if we
compute the OPE's of $\beta'$: \eqn\murmox{\beta'(z)\beta'(z')\sim
-{4\over (z-z')^2}\gamma^2(z')-{4\over
z-z'}\gamma\partial_{z'}\gamma(z').} As explained in \cdo, the
appropriate formula is \eqn\ucuc{\tilde\beta(z)=
-\beta\gamma^2(z)+2\partial_z\gamma(z).} $\tilde\beta$ has no
short distance singularity with itself and has the proper short
distance singularity with $\tilde\gamma$.

So this gives the appropriate description of a sheaf of CDO's that
is globally defined on the target space $\Bbb{CP}^1$ (but only
locally on the worldsheet $\Sigma$ of the conformal field theory,
since we defined it using a local complex parameter $z$). In this
example, one might think of $\tilde\beta$ as the ``correctly''
normal-ordered version of $-\beta\gamma^2$. In more complicated
examples, anomalies (obstructing existence of the theory) or free
parameters (moduli) arise in trying to find the right gluing.

\bigskip\noindent{\it Global Sections Of The Sheaf}

Having understood the sheaf $\hat {\cal A}$ of chiral  operators,
let us consider the global chiral algebra ${\cal A}$ of such
operators.  We recall that operators in ${\cal A}$ correspond to
elements of $H^i(\Bbb{CP}^1,\hat {\cal A})$.  As $\Bbb{CP}^1$ has
complex dimension 1, we have here  $i=0,1$. We write ${\cal A}^i$
for $H^i(\Bbb{CP}^1,\hat {\cal A})$.

First we consider ${\cal A}^0$, that is, the global sections of
$\hat{\cal A}$.  At dimension 0, we must consider functions of
$\gamma$ only. But a holomorphic function on $\Bbb{CP}^1$, to have
no poles anywhere, must be a constant, so the space of dimension 0
global sections is one-dimensional, generated by 1.

Typically, given a global section ${\cal O}$, we can make another
one of dimension one higher by differentiating, ${\cal O}\to
\partial_z{\cal O}$.  For the case of the identity operator, this
fails, as $\partial_z1=0$.

On the other hand, there are global sections of $\hat{\cal A}$ of
dimension 1.  There are no such sections of the form
$f(z)\partial_z\gamma$. Indeed, such an operator could be
transformed purely geometrically under \zino, by virtue of which
it would correspond to a holomorphic differential
$f(\gamma)d\gamma$ on $\Bbb{CP}^1$. But there are no such objects.

The remaining possibility is to find an operator that is linear in
$\beta$.  In fact, we right away see an example in \ucuc, as the
left hand side, $\tilde\beta$, is by definition regular in $U_2$,
while the right hand side, being polynomial in $\beta$, $\gamma$,
and their derivatives, is manifestly regular in $U_1$.  Their
being equal means that they represent a global section of
$\hat{\cal A}$ that we will call $J_+$:
\eqn\miklo{J_+=-\gamma^2\beta+2\partial\gamma =\tilde\beta.}

The construction is completely symmetric between $U_1$ and $U_2$,
with $\gamma\leftrightarrow \tilde\gamma$ and
$\beta\leftrightarrow\tilde\beta$, so a reciprocal formula gives
another dimension one global section $J_-$:
\eqn\inci{J_-(z)=\beta(z)=-\tilde\gamma^2\tilde \beta
+2\partial\tilde\gamma.} The skeptical reader can properly define
$\gamma^2\beta$ and similarly $\tilde\gamma^2\tilde\beta$ using
\ino\ and then verify that the gluing laws defining $\tilde\beta$
and $\tilde\gamma$ in terms of $\beta$ and $\gamma$ can be
inverted to solve for $\beta$ and $\gamma$ in terms of
$\tilde\beta $ and $\tilde\gamma$.

So $J_+$ and $J_-$ give us two dimension one sections of the sheaf
${\hat A}$.   Since these are global sections of a sheaf of chiral
vertex operators, we can construct more from their OPE's. There
are no singularities in the $J_+\cdot J_+$ or $J_-\cdot J_-$
operator products, but \eqn\igno{J_+(z)J_-(z')\sim {2J_3\over
z-z'}-{2\over (z-z')^2},} where $J_3$ is another global section of
dimension 1, \eqn\pigno{ J_3(z)=-\gamma\beta(z)} (which we again
define by normal-ordering).  For operator products involving
$J_3$, we get \eqn\invo{\eqalign{J_3(z)J_3(z') & \sim - {1\over
(z-z')^2}\cr J_3(z)J_+(z') & \sim {J_+(z')\over z-z'} \cr
     J_3(z)J_-(z') &\sim -{J_-(z')\over z-z'}.\cr}}

Taken together, the $J$'s generate a familiar chiral algebra --
the current algebra of $SL(2)$ at level $-2$, which here, as noted
in \cdo, appears in the Wakimoto free field representation
\ref\wak{M. Wakimoto, ``Fock Space Representations Of The Affine
Lie Algebra $A_1{}^{(1)}$, Commun. Math. Phys. {\bf 104} (1986)
609.}.  The space ${\cal A}^0$ of global sections of $\hat{\cal
A}$ is thus a module for this chiral algebra.

It is shown in \cdo\ to be an irreducible module, but one that has
unusual properties.  In general, for any $SL(2)$ current algebra
at any level $k\not= -2$, one can define a stress tensor
\eqn\uvvip{T(z)={:J_+J_-+J_3^2:\over k+2}.} For every $k\not=2$,
$T$ generates a Virasoro algebra.  If we want an operator that
makes sense at $k+2=0$, we can remove the factor of $1/(k+2)$ and
define \eqn\nuvvip{S(z) =:J_+J_-+J_3^2:.} Because
$S(z)=(k+2)T(z)$, it generates in its OPE's with any operator
$k+2$ times the transformation usually generated by the stress
tensor. If $k+2=0$, $S(z)$ generates no transformation at all --
it has no singularities in its OPE with any operator. Hence, in an
irreducible representation of current algebra, $S(z)$ can be
represented by a $c$-number, and might vanish. This fact is
important in conformal field theory approaches to the geometric
Langlands program (for reviews, see \frenkel). In that context, it
is important to consider a generalization in which $S(z)$ is set
to an arbitrary projective connection (or locally, a quadratic
differential) on $\Sigma$. That is what we get here if we carry
out the same construction using some other complex parameter on
$\Sigma$ instead of $z$. In fact, in this particular Wakimoto
module, it is true that $S(z)=0$.  The semiclassical approximation
to this statement can be verified immediately; with $J_-=\beta$,
$J_+=-\gamma^2\beta+\dots$, and $J_3=-\gamma\beta$, it is clear
that, ignoring quantum contractions, the $\beta^2$ terms in
$J_+J_-+J_3^2$ cancel.

As explained in \cdo, the space ${\cal A}^0$ of global sections of
$\hat{\cal A}$ is an irreducible module of $SL(2)$ current algebra
at level $-2$ that can be obtained from a free Verma module by
setting to zero $S$ and all of its derivatives. ${\cal A}^0$ has
the structure, roughly speaking, of a chiral algebra; it obeys all
the usual physical axioms of a chiral algebra, except the
existence of a stress tensor, and hence, reparameterization
invariance on the $z$-plane.

At one level, we have already explained why there is no stress
tensor: the usual definition \uvvip\ does not make sense at
$k=-2$.  But at another level, this may still appear perplexing.
The free $\beta\gamma$ system certainly has a stress tensor
\eqn\icop{T_f(z)=-:\beta\partial\gamma:(z)=-\lim_{z'\to
z}\left(\beta(z')\partial\gamma(z)+{1\over (z'-z)^2}\right).}
Likewise there is a free stress tensor of the
$\tilde\beta\tilde\gamma$ system, \eqn\nicop{\tilde
T_f(z)=-:\tilde \beta\partial\tilde\gamma:(z)=-\lim_{z'\to
z}\left(\tilde \beta(z')\partial\tilde\gamma(z)+{1\over
(z'-z)^2}\right).} Clearly $T_f$ is regular in $U_1$ and $\tilde
T_f$ is regular in $U_2$. The problem is that in $U_1\cap U_2$,
$T_f\not= \tilde T_f$, and there is no way to fix this (by adding
to $T_f$ a term regular in $U_1$ and to $\tilde T_f$ a term
regular in $U_2$).

If one inserts the definition of $\tilde\beta$ and $\tilde\gamma$
in the definition of $\tilde T_f$, a small computation shows that
\eqn\icvop{\tilde T_f-T_f=\partial\left({\partial\gamma\over
\gamma}\right).}

There is no way to fix this inconsistency without spoiling the
fact that $T_f$ and $\tilde T_f$ have the OPE's of stress tensors.
In fact, to preserve the OPE's $T_f\cdot \gamma$ and $\tilde
T_f\cdot \gamma$, we must add to $T_f$ or $\tilde T_f$ terms that
depend only on $\gamma$ or $\tilde \gamma$.  The right hand side
of \icvop\ is invariant under $\gamma\to\lambda\gamma$,
$\lambda\in \Bbb{C}^*$, and if it is possible to modify $T_f$ and
$\tilde T_f$ so as to agree on $U_1\cap U_2$, it can be done while
perserving this invariance. The only $\Bbb{C}^*$-invariant
operators of dimension two depending only on $\gamma$ are
$\partial^2\gamma/\gamma$ and $(\partial\gamma)^2/\gamma^2$. Any
linear combination of these has a pole at both $\gamma=0$ and
$\tilde\gamma=0$, so  these operators are of no help in removing
the anomaly.

So finally, although the free $\beta\gamma$ and
$\tilde\beta\tilde\gamma$ systems have stress tensors, there is no
stress tensor for the global chiral algebra ${\cal A}^0$ of
$\Bbb{CP}^1$.  This is a reflection of the fact that the $(0,2)$
model with target space $\Bbb{CP}^1$ is not conformally invariant
-- there is a non-zero one-loop beta function.  However, the
ability to see the failure of conformal invariance purely in terms
of holomorphic data \cdo\ is novel from the point of view of
physicists, which is why it has been expounded here in some
detail.  Moreover, the fact that conformal invariance is not
restored even for the chiral algebra, despite its holomorphy in
$z$, is somewhat surprising, at least for physicists.  Chiral
algebras without stress tensors, such as $SL(2)$ current algebra
at level $-2$, are important in the geometric Langlands program
\frenkel.

\nref\silv{E. Silverstein and E. Witten, ``Global $U(1)$ $R$
Symmetry And Conformal Invariance
Of $(0,2)$ Models,'' Phys. Lett. {\bf B328} (1994) 307, hep-th/9403054.}%
By contrast, $(0,2)$ models that are expected to flow to conformal
field theories in the infrared do typically have a stress tensor
in the chiral algebra, which hence is conformally invariant. This
has been seen in several examples of linear sigma models
\refs{\polyo,\silv}, and for nonlinear sigma models has been
proved, in the present context, when $c_1(X)=0$ \ngog.

Although the chiral algebra is not invariant under arbitrary
reparameterizations of $z$, it is invariant under affine
transformations $z\to az+b$, as these leave fixed the normal
ordering recipe.  In particular, from the scaling symmetry $z\to
az$, it follows that ${\cal A}^0$ and ${\cal A}^1$ are, in
perturbation theory, naturally graded by dimension, even though
there is no conformal invariance. Nonperturbatively, via
instantons, this grading by dimension is violated, unless
$c_1(X)=0$.

\bigskip\noindent{\it The First Cohomology}

Now we move on to investigate ${\cal A}^1=H^1(\Bbb{CP}^1,\hat{\cal
A})$.

There is no nonzero element of ${\cal A}^1$ of dimension zero.
Such an element would be represented as a function $f(\gamma)$,
with possible poles at $\gamma=0$ and $\gamma=\infty$, that cannot
be written as $f_1-f_2$, where $f_1$ is holomorphic in $U_1$ and
$f_2$ in $U_2$.  Such an $f$ does not exist, as
$H^1(\Bbb{CP}^1,{\cal O})=0$.

In dimension 1, ${\cal A}^1$ is one-dimensional, generated by an
object
$\theta=\partial\gamma/\gamma=-\partial\tilde\gamma/\tilde\gamma$
that we have essentially already encountered.  $\theta$ has a pole
at $\gamma=0$ and one at $\tilde\gamma=0$, and there is no way to
``split'' it as a difference of operators with only one pole. This
statement corresponds in ordinary geometry to the fact that the
differential $d\gamma/\gamma$, which is holomorphic in $U_1\cap
U_2$, generates $H^1(\Bbb{CP}^1,K)$, with $K$ the sheaf of
holomorphic differentials.  (If $d\gamma/\gamma$ could be
``split,'' the contour integral $\oint_Cd\gamma/\gamma$, for $C$ a
circle surrounding the pole at $\gamma=0$, would vanish.)

What happens in dimension 2?  At first it seems that we can
construct the dimension two operator $\partial\theta$ in ${\cal
A}^1$.  Certainly, classically, differentiating $\theta$ with
respect to $z$ does not help us in ``splitting'' it between $U_1$
and $U_2$.  But quantum mechanically, it is a different story. In
fact, this is what we get by reading \icvop\ backwards:
\eqn\hornof{\partial\theta=\tilde T_f-T_f.} Here $\tilde T_f$ is
holomorphic in $U_2$, and $T_f$ in $U_1$, so $\partial\theta$
vanishes as an element of  ${\cal A}^1$.

We can construct other elements of ${\cal A}^1$ by acting on
$\theta$ with the elements of ${\cal A}^0$.  Some relations are
obvious: $\partial\theta=0$, as we have just seen, and $S$ and all
its derivatives annihilate $\theta$, since $S$ actually vanishes
in ${\cal A}^0$.  It is shown in \cdo\ that these are the only
relations.  As a result, ${\cal A}^1$ is isomorphic to ${\cal
A}^0$, but shifted in dimension by 1.  This isomorphism maps the
dimension zero generator of ${\cal A}^0$, namely $1$, to the
dimension one generator of ${\cal A}^1$, namely $\theta$.   Of
course, such an isomorphism, shifting the dimension of the
operators, would not be compatible with conformal invariance.

The isomorphism between ${\cal A}^1$ and ${\cal A}^0$ is a
classical starting point for an important quantum phenomenon,
which will be discussed elsewhere.  ${\cal A}^1$ and ${\cal A}^0$
have been constructed to be annihilated in perturbation theory by
the differential $Q$ that we studied in sections 2 and 3.  But
nonperturbatively, $Q$ is corrected by instantons and the
corrected $Q$ is simply the isomorphism from ${\cal A}^1$ to
${\cal A}^0$.  As a result, in the exact quantum theory, the
cohomology of $Q$ in the space of local operators is identically
zero.  In particular, there is an instanton-induced relation
$\{Q,\theta\}\sim 1$; the fact that the identity operator is of
the form $\{Q,\dots\}$ implies that it acts trivially on the
${\cal A}$-module ${\cal V}$ given by the $Q$-cohomology of
quantum states (this module was briefly discussed at the end of
section 2.1).  So there are no supersymmetric states and
supersymmetry is spontaneously broken.

Finally, relation \hornof\ is an analog in \v{C}ech cohomology of
a formula that in conventional physical notation is familiar to
physicists and which in fact was briefly mentioned in \nologo:
\eqn\pologo{\partial_z(R_{i\bar j}\partial_z\phi^i\alpha^{\bar
j})=\{Q,T_{zz}\}.} Here, $R_{i\bar j}\partial_z\phi^i\alpha^{\bar
j}$ is the counterpart of $\theta$ in conventional physical
notation, and the relation \pologo\ can be read, just like
\hornof, in two ways.   It implies that the stress tensor $T$ is
not in the $Q$-cohomology, and that while $\theta$ does represent
an element of this cohomology, its derivative $\partial\theta$
vanishes in the $Q$-cohomology.

\subsec{The $p_1$ Anomaly}

Having illustrated in section 5.1 how the beta function can be
understood in the structure of the sheaf of chiral observables, we
now proceed to do the same for chiral anomalies. We consider here
the $p_1$ anomaly, and in section 5.3 we consider the
$c_1(X)c_1(\Sigma)$ anomaly.  We will recover results obtained in
the mathematical literature in \ngog\ (which uses the \v{C}ech
approach that we follow here) and also in \bd\ from a more
abstract viewpoint.

As a simple example of a complex manifold with $p_1\not= 0$, we
take $X=\Bbb{CP}^2$, which we endow with homogeneous coordinates
$\lambda_0,\lambda_1,\lambda_2$.  We cover $\Bbb{CP}^2$ with open
sets $U_i$, $i=0,\dots,2$, in which $\lambda_i\not=0$. We consider
the label $i$ to take values in $\Bbb{Z}/3\Bbb{Z}$, so $i=3$ is
equivalent to $i=0$.

 In each $U_i$, the sheaf of chiral operators
can be described by a free field theory, now with two
$\beta\gamma$ pairs since $\Bbb{CP}^2$ is of complex dimension
two.  We denote the spin zero fields in $U_i$ as $v^{[i]}$ and
$w^{[i]}$, and the spin one fields as $V^{[i]}$ and $W^{[i]}$. We
can take the spin zero fields to correspond to holomorphic
functions on $U_i$, as follows: \eqn\pokilp{\eqalign{
v^{[i]}&\leftrightarrow {\lambda_{i+1}\over \lambda_i}\cr w^{[i]}&
\leftrightarrow {\lambda_{i+2}\over \lambda_i}.\cr}} The action in
$U_i$  is \eqn\bomito{I^{[i]}={1\over 2\pi}\int
|d^2z|\left(V^{[i]}\bar\partial v^{[i]} +W^{[i]}\bar\partial
w^{[i]}\right).} The nontrivial OPE's of the fields appearing in
this action are \eqn\ifo{\eqalign{ V^{[i]}(z) & v^{[i]}(z')\sim
-{1\over z-z'}  \cr
                   W^{[i]}(z) & w^{[i]}(z')\sim -{1\over z-z'}.\cr}}
Other OPE's  are nonsingular.  To avoid cluttering the equations
too much, we also adopt a convention of writing simply $v,w,V$,
and $W$ as shorthand for $v^{[0]}$, $w^{[0]}$, $V^{0]}$, and
$W^{[0]}$.

To construct a global sheaf of chiral observables, we must find
gluing maps $R_i$ from operators in $U_i$ to operators in
$U_{i+1}$.  There is no problem in constructing any one $R_i$.
Moreover, as we have introduced the variables in a cyclically
symmetric way, the various $R_i$ can all be represented by
essentially the same formulas.   An anomaly appears because the
gluings are not compatible.  This will show up in the fact that
$R_2R_1R_0\not= 1$.

Of course, since the complex manifold $\Bbb{CP}^2$ does exist,
there is no inconsistency in the gluing at a geometrical level.
The difference $R_2R_1R_0-1$ is a nongeometrical symmetry of the
free field theory of $v,w,V$, and $W$, as described in section
3.4.

Such a symmetry is determined, we recall, by a closed holomorphic
two-form.  Consider a general system of $n$ conjugate
$\beta\gamma$ systems, with nontrivial OPE's
$\beta_i(z)\gamma^j(z')\sim -\delta_i^j/(z-z')$. Let $F={1\over
2}f_{ij}(\gamma)d\gamma^i\wedge d\gamma^j$ be a closed holomorphic
two-form.  Under the symmetry associated with $F$, the fields
transform as \eqn\bini{\eqalign{\gamma^j&\to\gamma^j\cr
                                \beta_i&\to
                               \beta_i'= \beta_i+f_{ij}\partial\gamma^j.\cr}}
In the spirit of section 3.4, one can justify this statement by
constructing locally a holomorphic one-form $A=A_id\gamma^i$ with
$dA=F$, and computing how the fields transform under the action of
the conserved charge $\oint A_i\partial\gamma^i$.  Alternatively,
one can simply check directly that the transformation \bini\
preserves the OPE's if $dF=0$. In general, the operator product of
$\beta'$ with itself gives
\eqn\hivno{\beta_i'(z')\beta'_j(z)\sim-{\partial\gamma^l\over
z'-z}\left(\partial_if_{kl}+\partial_kf_{li}+\partial_lf_{ik}\right).}
So the free field OPE's are preserved by the transformation \bini\
if and only if $dF=0$.

The anomaly will appear because $R_2R_1R_0-1$ will be a symmetry
of this nongeometrical kind, for some closed two-form $F$ that is
holomorphic in $U_0\cap U_1\cap U_2$, where the $R$'s are all
defined. Moreover, $F$ cannot be ``split'' as a sum of closed
two-forms $F_i$ that are holomorphic in $U_i\cap U_{i+1}$.  If
there were such a splitting, we would use it to correct the
individual $R_i$ so as to restore $R_2R_1R_0=1$.

The anomaly thus represents, as reviewed in section 3.5, an
element of $H^2(\Bbb{CP}^2,\Omega^{2,cl})$.  This group is
one-dimensional, and one can take a generator to be
\eqn\inon{F={dv\wedge dw\over vw},} which has poles when
$\lambda_0,\lambda_1,$ or $\lambda_2$ vanishes. The anomaly
therefore will appear in the fact that $R_2R_1R_0$, while leaving
$v$ and $w$ fixed, will transform $V$ and $W$ by
\eqn\jermy{\eqalign{V & \to V+ k{dw\over vw}\cr
                    W & \to W -k{dv\over vw}\cr}}
for some constant $k$.

\bigskip\noindent{\it The Computation}

Now that we know exactly what we are looking for, let us find it.

First, we find the transformation $R_0$ from $U_0$ to $U_1$. This
turns out to be \eqn\milop{\eqalign{v^{[1]} & = {w\over v}\cr
                                    w^{[1]} & = {1\over v}\cr
                                    V^{[1]} & = vW        \cr
                                    W^{[1]} & = -v^2V -vwW
                                    +{5\over 2}\partial v.\cr}}
The formulas for $v^{[1]}$ and $w^{[1]}$ are just the classical
changes of variable from $U_0$ to $U_1$, and likewise the terms in
$V^{[1]}$ and $W^{[1]}$ that are linear in $V$ and $W$ can be
found from classical geometry.  The last term in $W^{[1]}$ was
found as in section 5.1 to  ensure that the OPE's come out
correctly.

 As always, formulas such as those for $V^{[1]}$ and
$W^{[1]}$ are only meaningful if a precise recipe is given for
defining the operator products.  There are no ambiguities for
operators constructed only from $v$ and $w$ and their derivatives,
such as $w/v$ or $\partial w/v^2$.  We wlll interpret an operator
$f(v,w)V$  to mean \eqn\olko{f(v,w)V(z)=\lim_{z'\to
z}\left(f(v,w)(z')\cdot V(z)-{1\over z'-z}\partial_v
f(v,w)(z)\right),} and similarly for an operator $g(v,w)W$.  For
example, in \milop, we have \eqn\honn{\eqalign{vW(z)&=\lim_{z'\to
z}v(z')W(z)\cr
                           v^2V(z)&=\lim_{z'\to
                           z}\left(v^2(z')V(z)-{2v(z')\over
                           z'-z}\right).\cr}}
With this definition, the formulas in \milop\ have the correct
OPE's.  We use the same recipes for operator products for all
operators that appear presently, for example
\eqn\olko{f(v^{[i]},w^{[i]})V^{[i]}(z)=\lim_{z'\to
z}\left(f(v^{[i]},w^{[i]})(z')\cdot V^{[i]}(z)-{1\over
z'-z}\partial_{v^{[i]}} f(v^{[i]},w^{[i]})(z')\right),} for all
$i$.

Now, to obtain the transformation $R_1$ from $U_1$ to $U_2$, we
simply repeat this process: \eqn\umilop{\eqalign{v^{[2]} & =
{w^{[1]}\over v^{[1]}}\cr
                                    w^{[2]} & = {1\over v^{[1]}}\cr
                                    V^{[2]} & = v^{[1]}W^{[1]}        \cr
                                    W^{[2]} & = -(v^{[1]})^2V^{[1]}
                                    -v^{[1]}w^{[1]}W^{[1]}
                                    +{5\over 2}\partial v^{[1]}.\cr}}
Next, we substitute \milop\ into \umilop, so as to express
$v^{[2]},w^{[2]},\dots$ in terms of the original variables
$v,w,\dots,$
 and thereby get an explicit formula for the composition
$R_1R_0$. Here we have to be quite careful in the use of \olko. We
first express the operator products on the right hand side of
\umilop\ in a well-defined form, using \olko, to get a
well-defined formula for $v^{[2]},w^{[2]},\dots,$ in terms of
$v^{[1]},w^{[1]},\dots$, and then we substitute the expressions in
\milop\ to re-express those formulas in terms of $v,w,V,W$.  Upon
doing this, we obtain the following formulas: \eqn\kumilop{
\eqalign{ v^{[2]} & = {1\over w}\cr w^{[2]} & = {v\over w}\cr
V^{[2]}&=-vwV-w^2V+{3w\partial v\over 2v}+\partial w\cr
W^{[2]}&=wV+{3\partial w\over 2v}.\cr}} One can check these
formulas by verifying that the OPE's are correct.

The transformation $R_2$ from $U_2$ back to $U_3=U_0$ is, of
course, defined by the same formulas:
\eqn\upmilop{\eqalign{v^{[3]} & = {w^{[2]}\over v^{[2]}}\cr
                                    w^{[3 ]} & = {1\over v^{[2]}}\cr
                                    V^{[3]} & = v^{[2]}W^{[2]}        \cr
                                    W^{[3]} & = -(v^{[2]})^2V^{[2]}
                                    -v^{[2]}w^{[2]}W^{[2]}
                                    +{5\over 2}\partial v^{[2]}.\cr}}
Combining this with \kumilop, and again exercising care with the
definition of the operator products, we finally get an explicit
formula for the action of the composition $R_2R_1R_0$.  This
transformation acts on the field variables by \eqn\mombo{\eqalign{
v & \to v \cr w & \to w \cr V & \to V+{3\over 2}{\partial w\over
vw}\cr W& \to W-{3\over 2}{\partial v\over vw},\cr}}  exhibiting
the promised anomaly.

\subsec{The $c_1(\Sigma)c_1(X)$ Anomaly}

In a similar spirit, we can illustrate the $c_1(\Sigma)c_1(X)$
anomaly.  For this, we return to the example of section 5.1,
$X=\Bbb{CP}^1$.  But now, instead of working only locally on
$\Sigma$, as in section 5.1, we take $\Sigma=\Bbb{CP}^1$ and work
globally on $\Sigma$.

In section 5.1, we covered $X$ by two open sets $U_1$ and $U_2$,
respectively the complex $\gamma$-plane and $\tilde\gamma$-plane,
with $\tilde\gamma=1/\gamma$. Similarly, we can regard
$\Sigma=\Bbb{CP}^1$ as the complex $z$-plane glued to the complex
$y$-plane by the gluing map $y=1/z$.  We denote the $z$-plane as
$P_1$ and the $y$-plane as $P_2$.  A free $\beta\gamma$ system
defines a sheaf of chiral observables on $P_1$ or $P_2$ with
target $U_1$ or $U_2$.   As long as the target space is just $U_1$
or $U_2$, there is no problem in gluing together the theories
defined on $P_1$ and $P_2$, since the free $\beta\gamma$ system
makes sense on any Riemann surface. Similarly, as long as the
Riemann surface on which we define our theory is just $P_1$ or
$P_2$, we learned in section 5.1 how to glue together the theories
in which the target is $U_1$ or $U_2$ to make a theory with target
$\Bbb{CP}^1$. However, as discussed in general terms in section
3.5, an anomaly arises if we try to glue in both the $\Sigma$ and
the $X$ directions.

As usual, the anomaly involves a nongeometrical symmetry that acts
only on $\beta$. We briefly make some remarks on such symmetries
for the general case that $X$ has complex dimension $n$.
Nongeometrical symmetries, as we reviewed in section 3.2, are
determined by a closed holomorphic two-form $F$ on $X\times
\Sigma$.  If $F={1\over 2}f_{ij}(\gamma,z)d\gamma^i\wedge
d\gamma^j +C_i(\gamma,z)d\gamma^i\wedge dz$ is such a form, then
under the symmetry associated with $F$, the fields transform as
\eqn\ingo{\eqalign{\gamma^i& \to \gamma^i\cr
                   \beta_j & \to
                   \beta'_j=\beta_j+f_{ij}\partial\gamma^j+C_i.\cr}}
We get the OPE \eqn\pingo{\beta_i'(z')\beta_k'(z)\sim {1\over
z'-z}\left(-\partial\gamma^l(\partial_lf_{ik}+\partial_kf_{li}+\partial_if_{kl})
-(\partial_zf_{ik}+\partial_iC_k-\partial_kC_i)\right),} showing
that the transformation \ingo\ preserves the OPE's if and only if
$dF=0$.

The anomaly arises from an element of $H^2(X\times
\Sigma,\Omega^{2,cl})$ that appears as an inconsistency in gluing
together various local descriptions.  For $X=\Sigma=\Bbb{CP}^1$,
this cohomology group is one-dimensional.  A convenient generator
is the two-form \eqn\nbon{F={d\gamma\wedge dz\over \gamma \,z},}
which is holomorphic in all triple intersections of the open sets
$U_i\times P_j$.  Under the associated symmetry, the fields
transform as \eqn\ncono{\eqalign{\gamma & \to \gamma  \cr \beta
&\to \beta+{1\over \gamma z},\cr}} and this is the form that the
anomaly will take.

\bigskip\noindent{\it The Calculation}

We start with a free $\beta\gamma$ system, where $\gamma(z)$
describes a map from the $z$-plane to the $\gamma$-plane -- that
is, from the open set $P_1\subset \Sigma$ to the open set
$U_1\subset X$. In section 5.1, we showed how to map from $\gamma$
to $1/\gamma$, that is, from variables associated with $U_1\times
P_1$ to variables associated with  $U_2\times P_1$.  The gluing
map, which we will call $R$, is
\eqn\glumap{\eqalign{\tilde\gamma(z) & = {1\over \gamma(z)}\cr
                    \tilde \beta(z) & = \lim_{z'\to
                    z}\left(-\gamma^2(z')\beta(z)+{2\gamma(z')\over
                    z'-z}\right)+2\partial_z\gamma(z).\cr}}
Likewise, it is possible, while keeping the target space as $U_1$,
to glue together theories on which the Riemann surface is $P_1$ or
$P_2$.  In an elementary way, we can map a free $\beta\gamma$
system on the $z$-plane $P_1$ to a similar free field theory on
the $y$-plane $P_2$ (where $y=1/z$). We let $B$ and $\Gamma$ be
fields of dimension $(1,0)$ and $(0,0)$ on $P_2$, defined in terms
of $\beta$ and $\gamma$ by
\eqn\lumap{\eqalign{\Gamma(y)&=\gamma(1/y)\cr B(y)&=-{1\over
y^2}\beta(1/y).\cr}} This transformation, which we will call $Y$,
maps free field OPE's of $\beta$ and $\gamma$ to OPE's of the same
form for $B$ and $\Gamma$; the nontrivial OPE is
$\Gamma(y)B(y')\sim 1/(y-y')$.

By using $R$ to map from $U_1$ to $U_2$ and $Y$ to map from $P_1$
to $P_2$, we could -- if $R$ and $Y$ would commute -- glue
together four different free field descriptions associated with
$U_i\times P_j$ to make a theory that is global in both $X$ and
$\Sigma$. $R$ and $Y$ do commute in their action on $\gamma$.
Their combined operation on $\gamma$, in either order, gives
\eqn\unu{\hat\Gamma(y)={1\over\gamma(1/y)}.}
 But they do not commute  in their action on $\beta$.   That is where the anomaly comes
in.

 Let us write
$\tilde B$ for the result of applying first $R$ and then $Y$ -- in
other words, first mapping from $U_1$ to $U_2$ via \glumap, and
then from $P_1$ to $P_2$ via \lumap. The composition is easy to
write: \eqn\ononil{\tilde B(y)=-{1\over y^2}\left(\lim_{y'\to
y}\left(-\gamma^2(1/y')\beta(1/y)+{2\gamma(1/y')\over
1/y'-1/y}\right)+2\left.\partial_z\gamma\right|_{z=1/y}\right).}
And let us write $B^*$ for the result of reversing the order of
the two operations, applying first $Y$ and then $R$. Here we get
\eqn\koli{B^*(y)= \lim_{y'\to
y}\left(-\gamma^2(1/y')\left(-\beta(1/y)/y^2\right) +{2\over
y'-y}\gamma(1/y')\right)+2\partial_y\gamma(1/y).} When we subtract
these expressions, the $\gamma^2\beta$ terms trivially cancel, and
the $\partial\gamma$ terms cancel, given that $z=1/y$. The terms
linear in $\gamma$ do not cancel.  We get \eqn\inco{\tilde
B(y)-B^*(y)={2\gamma(1/y)\over y}={2\over y\hat\Gamma(y)},}
showing the form of the anomaly expected from \ncono.  Of course,
we get $y\hat\Gamma$ instead of $z\gamma$ in the denominator
because the equation \inco\ is written for fields on $U_2\times
P_2$.

\subsec{${\bf{S}}^1\times {\bf{S}}^3$ Revisited}

\def\XX{{\bf{S}}^1\times {\bf{S}}^3}
Finally, we will reexamine the ${\bf{S}}^1\times {\bf{S}}^3$
model, which we introduced in section 4.2.

\bigskip\noindent{\it The WZW Model}

First we make a few remarks on the WZW model of $\XX$, which as we
recall has $(0,2)$ supersymmetry \refs{\spindel,\rocek}, leading
to the possibility of constructing sheaves of CDO's on $\XX$.

This model is the tensor product of an $SU(2)$ WZW model at level
$k$, times a free field theory of $ {\bf{S}}^1$, times four free
right-moving (or in our conventions in this paper,
antiholomorphic) fermions.  These   fermions transform in the
adjoint representation of $SU(2)\times U(1)$.

After the twisting that is used, as in section 2.1, in defining
$Q$-cohomology and constructing a sheaf of CDO's, the fermions are
a pair of fermionic $\beta\gamma$ fields (called $\rho,\alpha$ in
section 2.1) with spins $1$ and $0$. A single such pair has left
and right central charges $(0,-2)$, so the fermions contribute
$(0,-4)$ to the left and right central charges of the system.  The
$SU(2)$ WZW model contributes $(3k/(k+2),3k/(k+2))$ to the central
charges, and the free theory of ${\bf{S}}^1$ contributes $(1,1)$.
The total central charges are hence $(3k/(k+2)+1,3k/(k+2)-3)$.
Passing from the physical theory to the $Q$-cohomology does not
change the difference of left and right central charges, which is
$c=4$. This will be the central charge of the stress tensor that
appears as a global section of the sheaf of CDO's.\foot{If one
replaces $\XX$ by $\Bbb{R}\times {\bf{S}}^3=\Bbb{C}^2-\{0\}$, it
is possible to add a linear dilaton coupling on $\Bbb{R}$ such
that the theory becomes a superconformal theory whose left and
right central charges (in the half-twisted version) are $(4,0)$.
In this description, the left-moving central charge is unchanged
in passing to the $Q$-cohomology, and remains at $c=4$.}

Similarly, we can anticipate the central charges of the current
algebra that will appear when we take global sections of the sheaf
of CDO's.  The underlying $SU(2)$ WZW model has an $SU(2)$-valued
field $g$, with symmetry $SU(2)_L\times SU(2)_R$  (actually
$(SU(2)_L\times SU(2)_R)/\Bbb{Z}_2$, where $\Bbb{Z}_2$ is the
common center of the two factors).  The symmetry acts by $g\to
agb^{-1}$, $a,b\in SU(2)$. In the WZW model, the $SU(2)_L$
symmetry is part of a holomorphic $SU(2)$ current algebra of level
$k$, while $SU(2)_R$ is part of an antiholomorphic $SU(2)$ current
algebra of level $k+2$.  Here, ``2'' is the contribution of the
right-moving fermions (real fermions in the adjoint representation
of $SU(2)$). The left and right central charges are thus $(k,0)$
for $SU(2)_L$ and $(0,k+2)$ for $SU(2)_R$.

The twisting of the four real fermions of the underlying $(0,2)$
model of $\XX$ to make a pair of fermionic $\beta\gamma$ or
$\rho\alpha$ systems explicitly breaks $SU(2)_R$ to its maximal
torus $U(1)_R$.  So the symmetry that survives for the
$Q$-cohomology or sheaf of CDO's is $(SU(2)_L\times
U(1)_R)/\Bbb{Z}_2=U(2)$.  The difference between left and right
central charges is unchanged in passing to the sheaf of CDO's, so
the level of the $SU(2)_L$ current algebra will be $k$ and that of
the $U(1)_R$ current algebra will be $-k-2$.  The only case in
which they are equal is $k=-1$, for which the levels are both
$-1$.   This is not really a physically sensible value for the WZW
model; physically sensible, unitary WZW models with convergent
path integrals are restricted to integer values of $k$ with $k\geq
0$.  However, in the sheaf of CDO's, which corresponds to a
perturbative treatment, $k$ is an arbitrary complex parameter, as
explained in sections 3 and 4.

In the sheaf of CDO's, the symmetries are automatically
complexified, so the symmetry we see will be at the Lie algebra
level $GL(2)$ rather than $U(2)$.    Moreover,  $U(1)_R$ (which
acts on the variables introduced momentarily by $v^i\to
e^{i\theta}v^i$) and the rotation of ${\bf{S}}^1$ (which acts by
$v^i\to e^{\chi}v^i$ with real $\chi$) combine together to
generate the center of $GL(2)$.  (At the Lie algebra level, the
center is $GL(1)$, but the group structure is really that of an
elliptic curve or of $GL(1)/\Bbb{Z}=U(1)\times U(1)$.) The
rotation of ${\bf{S}}^1$ corresponds to a $U(1)$ current algebra
with equal left and right central charges, so it does not affect
the above discussion.

\bigskip\noindent{\it Constructing A Sheaf Of CDO's}

Now we come to our main task, which is to construct a family of
sheaves of CDO's over $\XX$.  First we construct a simple special
case, and then we show how to introduce a variable parameter.

We will regard $\XX$ as $(\Bbb{C}^2-\{0\})/\Bbb{Z}$, where
$\Bbb{C}^2$ has coordinates $v^1,$ $v^2$, and $\{0\}$ is the
origin in $\Bbb{C}^2$ (the point $v^1=v^2=0$) which should be
removed before dividing by $\Bbb{Z}$. Also, $\Bbb{Z}$ acts by
$v^i\to \gamma^n v^i$, where $\gamma$ is a nonzero complex number
of modulus less than 1.  $\gamma$ is a modulus of $\XX$ that we
will keep fixed.

To construct the simplest sheaf of CDO's with target $\XX$, we
simply promote the $v^i$ to free fields of spin 0, with conjugate
spin 1 fields $V_i$, and free action \eqn\polko{I={1\over
2\pi}\int|d^2z|\left(V_1\bar\partial v^1+V_2\bar\partial
v^2\right).  } The nontrivial OPE's are as usual
$V_i(z)v^j(z')\sim -\delta^i_j/(z-z')$.  We allow only operators
that are invariant under $v^i\to \gamma v^i$, $V_i\to
\gamma^{-1}V_i$.

One operator that possesses this invariance is the stress tensor
\eqn\inco{ T_{zz}=\sum_iV_i\partial v^i.} Hence, the chiral
algebra of this theory is conformally invariant, in contrast to
the chiral algebra of $\Bbb{CP}^1$.  This reflects the conformal
invariance of the underlying $(0,2)$ model with target $\XX$. A
bosonic $\beta\gamma$ system of spins 0 and 1 has $c=2$, so the
stress tensor $T$ has $c=4$, in agreement with what we expected
from the underlying WZW model.

The chiral algebra of $\XX$ also contains the dimension $1$
currents $J^i_j=-V_jv^i$.  These obey the OPE's
\eqn\murgo{J^i_j(z)J^m_l(z')\sim -{\delta^i_l\delta^m_j\over
(z-z')^2}+{\delta_j^mJ^i_l-\delta^i_lJ^m_j\over z-z'}.} This is a
$GL(2)$ current algebra at level $-1$.

In what follows, it will not be possible to maintain manifest
$GL(2)$ symmetry, and it will be convenient to pick a basis in the
current algebra. The $SL(2)$ subgroup is generated by $J_3=-\half
(V_1v^1-V_2v^2)$, $J_+=-V_2v^1$, $J_-=-V_1v^2$, with nontrivial
OPE's \eqn\ploko{\eqalign{J_3(z)J_3(z')&\sim -{1\over 2}{1\over
(z-z')^2}\cr J_3(z)J_{\pm}(z')&\sim \pm {J_{\pm}(z')\over z-z'}\cr
J_+(z)J_-(z')&\sim -{1\over (z-z')^2}+{2J_3(z')\over z-z'}.\cr}}
Here we recognize the $SL(2)$ current algebra at level $-1$.  The
center of $GL(2)$, which is of course a copy of $GL(1)$, is
generated by $K=-{1\over 2}\left(V_1v^1+V_2v^2\right)$, with
\eqn\milko{K(z)K(z')\sim -{1\over 2}{1\over (z-z')^2}.}

\bigskip\noindent{\it The Modulus Of The CDO}

Now we are going to generalize the CDO of $\XX$ that was
constructed above, introducing a parameter associated with
$H^1(\XX,\Omega^{2,cl})\cong \Bbb{C}$.

To do this, we first make a cover of $\XX$ by two open sets $U_1$
and $U_2$, where $U_1$ is characterized by the condition
$v^1\not=0$, and $U_2$ by $v^2\not= 0$.  In fact, this is not a
``good cover,'' as $U_1$ and $U_2$ are topologically complicated
(each is isomorphic to $\Bbb{C}\times E$, where $E$ is an elliptic
curve).  As a result, in general, we are not guaranteed that an
arbitrary cohomology class can be represented by a \v{C}ech
cocycle with respect to this cover.  In the case at hand, however,
we have on $U_1\cap U_2$ a holomorphic section of $\Omega^{2,cl}$,
namely \eqn\ubuc{F={dv^1\wedge dv^2\over v^1 v^2}.} $F$ cannot be
``split'' as the difference of a form holomorphic in $U_1$ and one
holomorphic in $U_2$, so it represents an element of
$H^1(\XX,\Omega^{2,cl})$.

 From \bini, we know that the symmetry associated with $F$ generates
 the following transformation: \eqn\ilvo{\eqalign{v^1&\to v^1\cr v^2 &\to v^2\cr V_1&\to
V_1+t{\partial v^2\over v^1v^2}\cr V_2&\to V_2-t{\partial v^1\over
v^1v^2}.\cr}} Here $t$ is a complex parameter, which will turn out
to be related to $k$ of the WZW model. We get a family of CDO's,
parameterized by $t$,  by declaring that  the fields undergo this
gluing in going from $U_1$ to $U_2$.

Let us determine how some important operators behave under this
deformation. The stress tensor $T=V_1\partial v^1+V_2\partial v^2$
is invariant. So the deformed theory, for any $t$, has a stress
tensor of $c=4$.  This is in accord with the fact that the WZW
model is conformally invariant for all $k$ and that the difference
of its left and right central charges is always 4.

Next, let us consider the $GL(1)$ current, which at $t=0$ was
defined as $K=-\half(V_1v^1+V_2v^2)$.  Under \ilvo, we have
\eqn\jilvo{K\to K-{t\over 2}\left({\partial v_2\over
v_2}-{\partial v_1\over v_1}\right).}  In contrast to what one
might guess from our previous examples, the shift in $K$ under
this transformation is {\it not} an anomaly that spoils existence
of $K$ at $t\not=0$. The reason is that this shift can be split as
a difference between a term (namely $t\,\partial v^1/2v^1$) that
is holomorphic in $U_1$ and a term (namely $t\,\partial v^2/2v^2$)
that is holomorphic in $U_2$. As a result, we can modify $K$ to
get a $GL(1)$ current generator that is holomorphic in both $U_1$
and $U_2$.  In $U_1$, the current is \eqn\ncon{K^{[1]}=-{1\over
2}\left(V_1v^1+V_2v^2\right)-{t\over 2}{\partial v^1\over v^1},}
while in $U_2$ it is represented by \eqn\bcon{K^{[2]}=-{1\over
2}\left(V_1v^1+V_2v^2\right)-{t\over 2}{\partial v^2\over v^2}.}
$K^{[1]}$ is holomorphic in $U_1$, and transforms under \ilvo\
into $K^{[2]}$, which is holomorphic in $U_2$.  So, for any $t$,
the sheaf $\hat{\cal A}$ of chiral operators has a global section
$K$ that is represented in $U_1$ by $K^{[1]}$ and in $U_2$ by
$K^{[2]}$.

Now we can calculate the OPE singularity of $K$ for any $t$:
\eqn\kini{K(z)K(z')\sim - {1+t\over 2} {1\over (z-z')^2}.} To
calculate this, we either work in $U_1$, setting $K=K^{[1]}$ and
computing the OPE, or we work in $U_2$, setting $K=K^{[2]}$ and
computing the OPE.  The answer comes out the same either way,
since the transformation \ilvo\ is an automorphism of the CFT.
Thus, the level of the $GL(1)$ current algebra is $-t-1$.

Similarly, we can work out the transformation of the $SL(2)$
currents under \ilvo.  The currents as defined at $t=0$, namely
$J_3=-\half(V_1v^1-V_2v^2)$, $J_+=-V_2v^1$, $J_-=-V_1v^2$,
transform as \eqn\ivo{\eqalign{J_3 & \to J_3-{t\over
2}\left({\partial v^1\over v^1}+{\partial v^2\over v^2}\right)\cr
J_+& \to J_+ +t{\partial v^1\over v^2}\cr J_-&\to J_- -t{\partial
v^2\over v^1}.\cr}} The shifts in each current can be ``split'' as
a difference of terms holomorphic in $U_1$ and $U_2$.  So the
currents can be defined at $t\not=0$, but  receive $t$-dependent
terms. The corrected currents are \eqn\bogus{J_3=
\cases{-\half\left(V_1v^1-V_2v^2\right)+{t\,\partial v^1/ 2v^1}\cr
-\half\left(V_1v^1-V_2v^2\right)-{t\,\partial v^2/ 2v^2}\cr}}
along with \eqn\nogus{J_+ =\cases{-V_2v^1\cr -V_2v^1+t\,{\partial
v^1/ v^2}\cr}} and \eqn\ogus{J_-=\cases{-V_1v^2+t\,\partial
v^2/v^1\cr -V_1v^2. }} In each case, the upper expression holds in
$U_1 $ and the lower expression holds in $U_2$.

We now can compute the OPE's of these operators, working in either
$U_1$ or $U_2$.  We get an $SL(2)$ current algebra at level $t-1$:
\eqn\ploko{\eqalign{J_3(z)J_3(z')&\sim {t-1\over 2}{1\over
(z-z')^2}\cr J_3(z)J_{\pm}(z')&\sim \pm {J_{\pm}(z')\over z-z'}\cr
J_+(z)J_-(z')&\sim {t-1\over (z-z')^2}+{2J_3(z')\over z-z'}.\cr}}

The $SL(2)$ and $GL(1)$ current algebras thus have levels $t-1$
and $-t-1$, in agreement with expectations from the WZW model if
the WZW level $k$ is related to the CDO parameter $t$ by $k=t-1$.
We will not attempt an {\it a priori} explanation of this
relationship.

The $Q$-cohomology of $\XX$ has no instanton corrections.  For any
target space $X$, such corrections (because they are local on the
Riemann surface $\Sigma$, though global in $X$) come only from
holomorphic curves in $X$ of genus zero.  There are no such curves
in $\XX$.

\bigskip
This work was supported in part by NSF Grant PHY-0070928.

 \listrefs
\end